\documentclass[fleqn,usenatbib]{mnras}

\usepackage{newtxtext,newtxmath}

\usepackage[T1]{fontenc}
\DeclareRobustCommand{\VAN}[3]{#2}
\let\VANthebibliography\thebibliography
\def\thebibliography{\DeclareRobustCommand{\VAN}[3]{##3}\VANthebibliography}

\usepackage{graphicx}	
\usepackage{amsmath}	
\usepackage[flushleft]{threeparttable}	
\usepackage{color}
\usepackage{booktabs}
\usepackage{mathtools, cuted}
\usepackage{stfloats}
\usepackage{acronym}
\usepackage{multirow}
\usepackage{float}
\usepackage{subfig}
\usepackage{soul}
\usepackage{makecell}
\usepackage{colortbl}
\usepackage{bbm}
\usepackage[dvipsnames]{xcolor}
\usepackage{pdfpages}
\usepackage{tabularx}
\usepackage{siunitx}
\usepackage{graphicx}

\newcommand{\rf}{RF}
\newcommand{\flex}{FlexCoDE}
\newcommand{\bnn}{BMDN}
\newcommand\given[1][]{\:#1\vert\:}
\newcommand{\dr}{DR4}

\newcommand{\RFrmse}{0.423}
\newcommand{\RFnmad}{0.100}
\newcommand{\RFbias}{-0.002}
\newcommand{\RFnquinze}{0.225}
\newcommand{\RFntrinta}{0.068}

\newcommand{\RFnbrmse}{0.408}
\newcommand{\RFnbnmad}{0.090}
\newcommand{\RFnbbias}{0.003}
\newcommand{\RFnbnquinze}{0.220}
\newcommand{\RFnbntrinta}{0.066}

\newcommand{\FLEXrmse}{0.477}
\newcommand{\FLEXnmad}{0.084}
\newcommand{\FLEXbias}{0.044}
\newcommand{\FLEXnquinze}{0.222}
\newcommand{\FLEXntrinta}{0.084}

\newcommand{\FLEXnbrmse}{0.455}
\newcommand{\FLEXnbnmad}{\textbf{0.039}}
\newcommand{\FLEXnbbias}{0.016}
\newcommand{\FLEXnbnquinze}{0.209}
\newcommand{\FLEXnbntrinta}{0.078}

\newcommand{\BNNrmse}{0.458}
\newcommand{\BNNnmad}{0.083}
\newcommand{\BNNbias}{0.025}
\newcommand{\BNNnquinze}{0.212}
\newcommand{\BNNntrinta}{0.074}

\newcommand{\BNNnbrmse}{0.427}
\newcommand{\BNNnbnmad}{0.048}
\newcommand{\BNNnbbias}{$\mathbf{10^{-4}}$}
\newcommand{\BNNnbnquinze}{\textbf{0.187}}
\newcommand{\BNNnbntrinta}{0.068}

\newcommand{\AVErmse}{0.420}
\newcommand{\AVEnmad}{0.086}
\newcommand{\AVEbias}{0.023}
\newcommand{\AVEnquinze}{0.204}
\newcommand{\AVEntrinta}{0.063}

\newcommand{\AVEnbrmse}{\textbf{0.392}}
\newcommand{\AVEnbnmad}{0.058}
\newcommand{\AVEnbbias}{0.006}
\newcommand{\AVEnbnquinze}{0.188}
\newcommand{\AVEnbntrinta}{\textbf{0.058}}

\title[The Quasar Catalogue for S-PLUS \dr{}]{The Quasar Catalogue for S-PLUS \dr{} (QuCatS) and the estimation of photometric redshifts}

\author[L. Nakazono \& R. R. Valen\c{c}a et al.]{L. Nakazono$^{1,2}$\thanks{Email: lilianne.nakazono@usp.br}, R. R. Valen\c{c}a$^{1}$, G. Soares$^{3}$,  R. Izbicki$^{3}$, \v{Z}. Ivezi\'{c}$^{2}$,  E.V. R. Lima$^{1}$, \\ 
\newauthor{N. S. T. Hirata$^{4}$, L. Sodr\'{e} Jr.$^{1}$, R. Overzier$^{5}$, F. {Almeida-Fernandes}$^{1, 6}$, G. B. Oliveira Schwarz$^{7}$,} \\
\newauthor{W. Schoenell$^{8}$, A. Kanaan$^{9}$, T. Ribeiro$^{10}$, C. Mendes de Oliveira$^{1}$} 
\\
$^{1}$ Universidade de S\~ao Paulo, Instituto de Astronomia, Geof\'isica e Ci\^encias Atmosf\'ericas, Departamento de Astronomia, SP 05508-090, S\~ao Paulo, Brazil\\
$^{2}$ Department of Astronomy and DiRAC Institute, University of Washington, Box 351580, Seattle, WA 98195, USA \\  
$^{3}$ Universidade Federal de S\~{a}o Carlos, Departamento de Estat\'istica, S\~{a}o Carlos, 13565-905, SP, Brazil \\  
$^{4}$ Universidade de S\~ao Paulo, Instituto de Matem\'{a}tica e Estat\'{i}stica, Departamento de Ci\^{e}ncia da Computa\c{c}\~{a}o, SP 05508-090, S\~ao Paulo, Brazil\\ 
$^{5}$ Observat\'orio Nacional / MCTIC, Rua General Jos\'e Cristino 77, Rio de Janeiro, RJ, 20921-400, Brazil \\
$^{6}$ NSF’s NOIRLab, 950 N. Cherry Ave., Tucson, AZ 85719, USA\\
$^{7}$ Universidade Presbiteriana Mackenzie, R. da Consolação, 930 - Consolação, São Paulo, Brazil\\
$^{8}$ GMTO Corporation 465 N. Halstead Street, Suite 250 Pasadena, CA 91107 \\
$^{9}$
Departamento de F\'isica, Universidade Federal de Santa Catarina, Florian\'opolis, SC, 88040-900, Brazil \\
$^{10}$ Rubin Observatory Project Office, 950 N. Cherry Ave., Tucson, AZ 85719, USA \\
}

\date{Accepted XXX. Received YYY; in original form ZZZ}
\pubyear{2023}

\begin{document}
\label{firstpage}
\pagerange{\pageref{firstpage}--\pageref{lastpage}}
\maketitle

\begin{abstract}
The advent of massive broad-band photometric surveys enabled photometric redshift estimates for unprecedented numbers of galaxies and quasars. These estimates can be improved using better algorithms or by obtaining complementary data such as narrow-band photometry, and broad-band photometry over an extended wavelength range. We investigate the impact of both approaches on photometric redshifts for quasars using data from Southern Photometric Local Universe Survey (S-PLUS) \dr, Galaxy Evolution Explorer (GALEX) DR6/7, and the unWISE catalog for the Wide-field Infrared Survey Explorer (WISE) in three machine learning methods: Random Forest, Flexible Conditional Density Estimation (FlexCoDE), and Bayesian Mixture Density Network (BMDN). Including narrow-band photometry improves the root-mean-square error by 11\% in comparison to a model trained with only broad-band photometry. Narrow-band information only provided an improvement of 3.8\% when GALEX and WISE colours were included. Thus narrow bands play a more important role for objects that do not have GALEX or WISE counterparts, which  respectively makes 92\% and 25\% of S-PLUS data considered here. Nevertheless, the inclusion of narrow-band information provided better estimates of the probability density functions obtained with \flex{} and \bnn{} . We publicly release a value-added catalogue of photometrically selected quasars with the photo-\textit{z} predictions from all methods studied here. The catalogue provided with this work covers the S-PLUS DR4 area ($\sim3000 \text{deg}^2$), containing 645\,980, 244\,912, 144\,991 sources with the probability of being a quasar higher than, 80\%, 90\%, 95\% up to $r<21.3$ and good photometry quality in the detection image. More quasar candidates can be retrieved from the S-PLUS data base by considering less restrictive selection criteria.
\end{abstract}



\section{Introduction}
\label{sec:introduction}

Quasars are known to be the most energetic objects in the Universe (\citealt{2013ARA&A..51..511K}). 
Since the first quasar discovery by \cite{1963Natur.197.1040S}, the number of confirmed quasars increased to almost a million. Among many surveys that have contributed to the discovery of new quasars, the Sloan Digital Sky Survey (SDSS; \citealp{2000AJ....120.1579Y}) has provided the largest number of confirmed quasars with the SDSS-I/II/III/IV phases, including the extended Baryon Oscillation Spectroscopic Survey (eBOSS; \citealt{2016AJ....151...44D}). The SDSS DR16 quasar catalogue from DR16 (DR16Q; \citealt{2020ApJS..250....8L}) includes objects from all SDSS programmes and contains 920\,110 spectroscopic observations of 750\,414 quasars. 
The SDSS success in finding quasars is due to multiple efficient techniques for candidate quasar selection and redshift estimation based on optical and mid-infrared colours (see \citealt{2015ApJS..221...27M} for an overview). 

Information needed to select candidate quasars for follow-up spectroscopic observations is typically provided 
by photometric sky surveys. Furthermore, if adequate colours are available, quasars redshifts can be estimated 
using photometry alone, thus bypassing the need for expensive spectroscopic observations. 
Techniques to estimate redshifts using photometric observations (photometric redshifts or photo-\textit{z}s) are 
well studied in the literature, but they remain an active research topic.
There are two main approaches to photo-\textit{z}s: SED (spectral energy distribution) template fitting (e.g. \citealt{2009ApJ...690.1250S}, \citealt{2004MNRAS.353..654B}) and empirical methods based on training samples (e.g. \citealt{2010MNRAS.406.1583W}, \citealt{2013ApJ...772..140B}, \citealt{2021A&A...649A..81N}, \citealt{2022arXiv220608989Y}), each with their own pros and cons. 
In general, machine learning as an empirical approach provides more accurate photo-\textit{z} predictions than SED template fitting (e.g. \citealt{2020MNRAS.499.1587S}). 
However, the robust performance of such methods is limited to the parameter space of the training set, while
SED template fitting has no such limitations (see \citealt{2021FrASS...8...70B} for further discussion).

In this work, we analyse the performance of photometric redshift estimates based on 12-band photometric data from the Southern Photometric Local Universe Survey (S-PLUS; \citealt{2019MNRAS.489..241M}). 
S-PLUS is a large-area sky survey that will cover $\sim$9300 deg$^2$ of the Southern Sky with the same optical system used in the Javalambre Photometric Local Universe Survey (J-PLUS; \citealt{2019A&A...622A.176C}). 
This optical system consists of seven narrow bands and five $ugriz$ broad bands ($griz$ are similar to the SDSS bands;  \citealt{1996AJ....111.1748F}), providing effectively low-resolution spectra 
that are expected to improve photometric redshift estimates compared to those based on broad-band photometry alone.
We also extend S-PLUS data with photometry from Galaxy Evolution Explorer (GALEX; \citealt{2005ApJ...619L...1M}) and Wide-field Infrared Survey Explorer (WISE; \citealt{2010AJ....140.1868W}) surveys to study how photo-\textit{z} estimates improve due to an extended wavelength range compared to improvements due to the addition of narrow-band photometry. 
We focus here on photo-\textit{z} techniques based on machine learning.

We tested three different and independent machine learning algorithms: Random Forest, Flexible Non-parametric Conditional Density Estimation (FlexCoDE), and Bayesian Mixture Density Network. 
We discuss the advantages and disadvantages of each method, and in particular assess the importance of the narrow-band photometry for the photo-\textit{z} regression. Using three independent methods allows us to check for consistency in the interpretation of our results.  We provide a value-added catalogue of quasar candidates for spectroscopic follow-up with their assigned probabilities of being a quasar (obtained from the star/quasar/galaxy classification by \citealt{2021MNRAS.507.5847N}) and their estimated photometric redshifts from this work. Estimates from all three methods are provided separately in the catalogue.

This paper is organised as follows: data sets used to estimate photo-\textit{z}s are described in Section \ref{sec:surveys} 
and photo-\textit{z} methods are described in Section \ref{sec:methods}. We present our results in Section \ref{sec:results} and how to select quasar candidates in Section \ref{sec:vac}. Finally, our main findings are discussed in Section \ref{sec:discussion}. 

\raggedbottom

\section{Photometric Surveys and Datasets}
\label{sec:surveys}

In this section, we describe three photometric surveys, S-PLUS, WISE, and GALEX, that provided photometry for photo-\textit{z} estimation. In addition, a reliable sample of known quasars with spectroscopic redshifts is needed for training supervised machine learning methods and we used the SDSS DR16Q quasar catalogue \citep{2020ApJS..250....8L}.  In Table \ref{tab:mag}, we list the bands used in this work together with their effective wavelength, 
width and magnitude depth.

\begin{table}
\centering
\caption{Optical filter system characterization of GALEX 
($5\sigma$; \citealt{2007ApJS..173..682M}), S-PLUS (peak of the magnitude distribution at $\text{S/N}>3$; \citealt{2022MNRAS.511.4590A}) and WISE (50\% completeness at $\text{S/N} > 5$; \citealt{2019ApJS..240...30S}). The effective wavelengths and widths are given in Angstroms, and the depths are given in AB magnitudes.}
\label{tab:mag}

\begin{tabular}{ccccc} 
\toprule
\textbf{Survey} & \textbf{Band} & \textbf{Effective Wavelength} & \textbf{Width} & \textbf{Depth}  \\ 

\midrule
\multirow{2}{*}{GALEX}  & $FUV$   & 1528  & 442   & 19.90 \\
                                 & $NUV$   & 2310  & 1060  & 20.80 \\

\midrule
\multirow{5}{*}{\begin{tabular}[c]{@{}c@{}}S-PLUS \\  (broad bands)\end{tabular}}  & $u$     & 3536  & 352   & 21.0 \\
                                 & $g$     & 4751  & 1545  & 21.3 \\
                                 & $r$     & 6258  & 1465  & 21.3 \\
                                 & $i$     & 7690  & 1506  & 20.9 \\
                                 & $z$     & 8831  & 1182  & 20.1 \\ \midrule

\multirow{7}{*}{\begin{tabular}[c]{@{}c@{}}S-PLUS \\ (narrow bands)\end{tabular}} & J0378 & 3770  & 151   & 20.4 \\
                                 & $J0395$ & 3940  & 103   & 19.9 \\
                                 & $J0410$ & 4094  & 201   & 20.0 \\
                                 & $J0430$ & 4292  & 201   & 20.0 \\
                                 & $J0515$ & 5133  & 207   & 20.2 \\
                                 & $J0660$ & 6614  & 147   & 21.1 \\
                                 & $J0861$ & 8611  & 408   & 19.9 \\ 
\midrule
\multirow{2}{*}{WISE}   & $W1$    & 34000 & 6600  & 20.72 \\
                                 & $W2$    & 46000 & 10400 & 19.97 \\
\bottomrule
\end{tabular}
\end{table}

\subsection{The Southern Photometric Local Universe Survey (S-PLUS)}

The S-PLUS survey is an ongoing survey that, by the end, will cover $\sim$9300 deg$^2$ of the Southern Sky with the T80-South telescope located at Cerro Tololo Inter-American Observatory (CTIO), Chile. The S-PLUS optical filter system includes five broad-bands: $u$, $g$, $r$, $i$, $z$ (similar to the SDSS filter system, except for the $u$ band) and seven narrow-band filters: $J0378$, $J0395$, $J0410$, $J0430$, $J0515$, $J0660$, and $J0861$, centred on [OII], Ca H+K, H$\gamma$, G-band, Mgb triplet, H$\alpha$ and Ca triplet features, respectively \citep{2012SPIE.8450E..3SM}.  

We use aperture fluxes computed within a circular diameter of 3 arcsec and corrected for the fraction of the flux falling outside this limit (see \citealt{2022MNRAS.511.4590A} for more details). The procedure for data reduction and calibration of S-PLUS Data Release 4 (DR4), used in this work, is further detailed in \citealt{herpich2024fourthsplusdatarelease}. The S-PLUS \dr{} comprises $\sim$3000 deg$^2$, including the area from past S-PLUS data releases.
There are 645 980, 244 912, 144 991 sources photometrically classified as quasars with 80\%, 90\%, and 95\% probabilities in S-PLUS with $r<21.3$ and good photometry flag in the detection image (SEX\_FLAGS\_DET $=0$). We corrected the data for reddening using dust maps from \cite{1998ApJ...500..525S} and the extinction law from \cite{1989ApJ...345..245C}

\begin{figure}
\centering
    \subfloat{\includegraphics[width=0.47\textwidth]{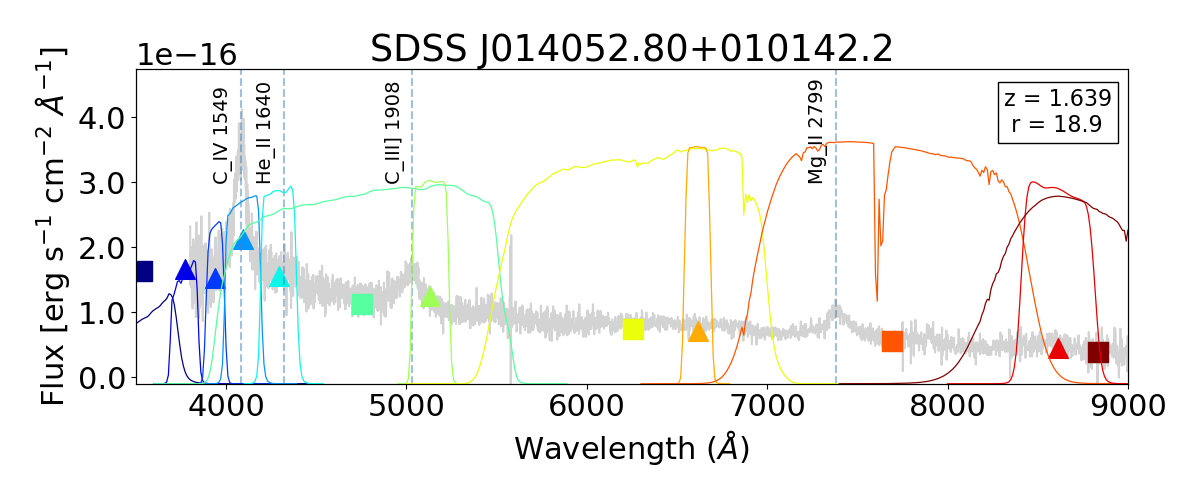}}
    
    \subfloat{\includegraphics[width=0.47\textwidth]{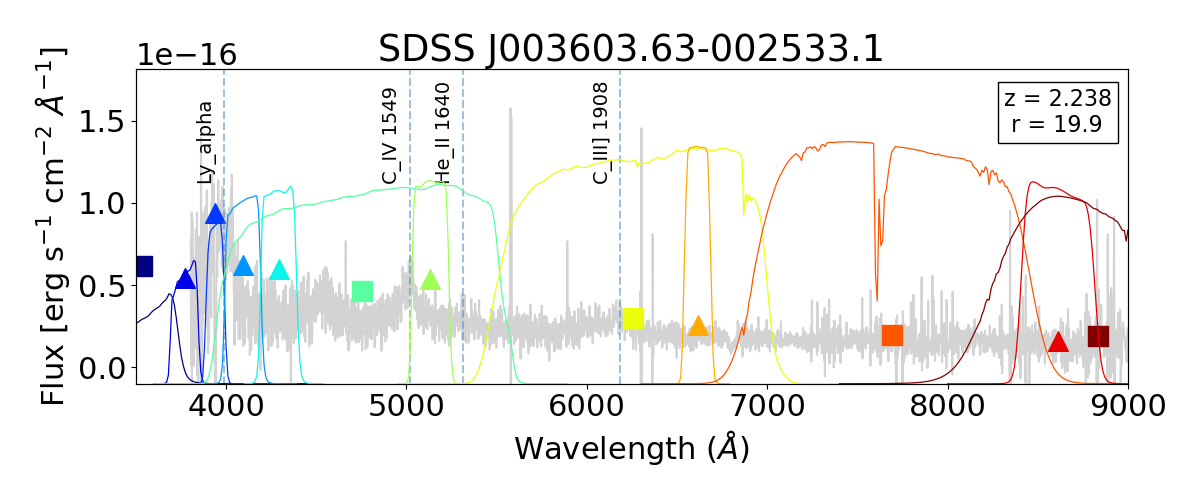}}
        
    \subfloat{\includegraphics[width=0.47\textwidth]{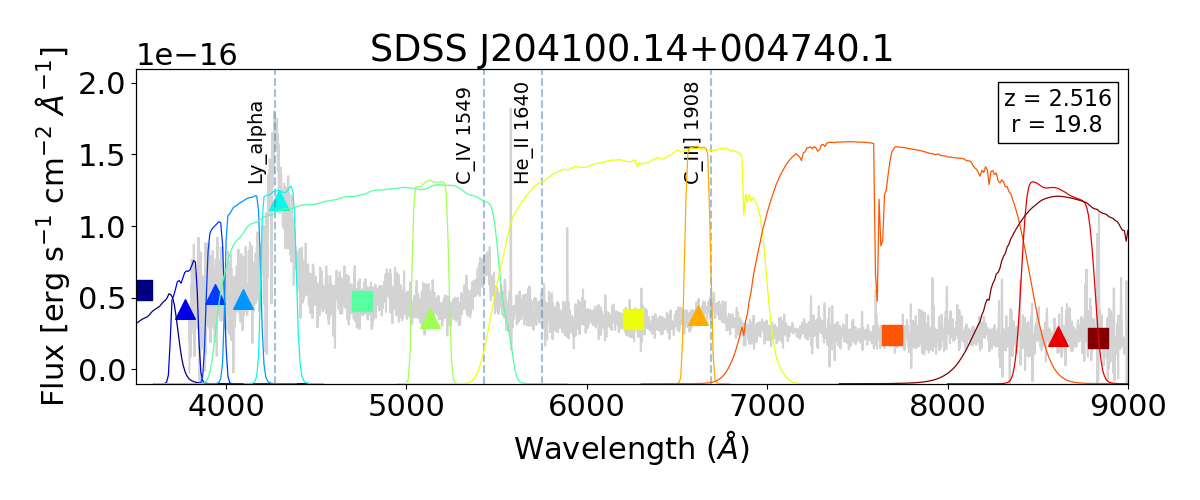}}

    \subfloat{\includegraphics[width=0.47\textwidth]{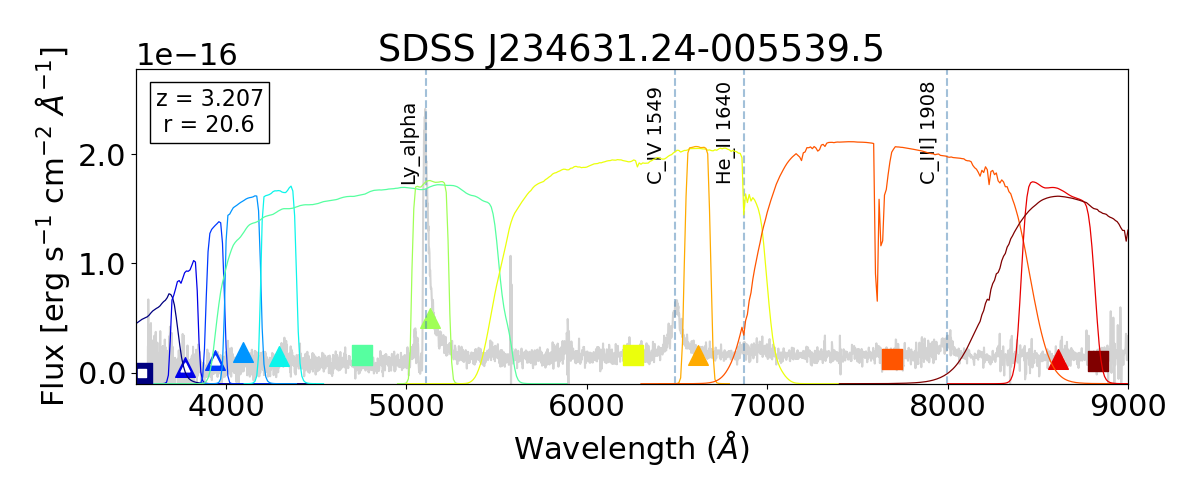}}
            
    \subfloat{\includegraphics[width=0.47\textwidth]{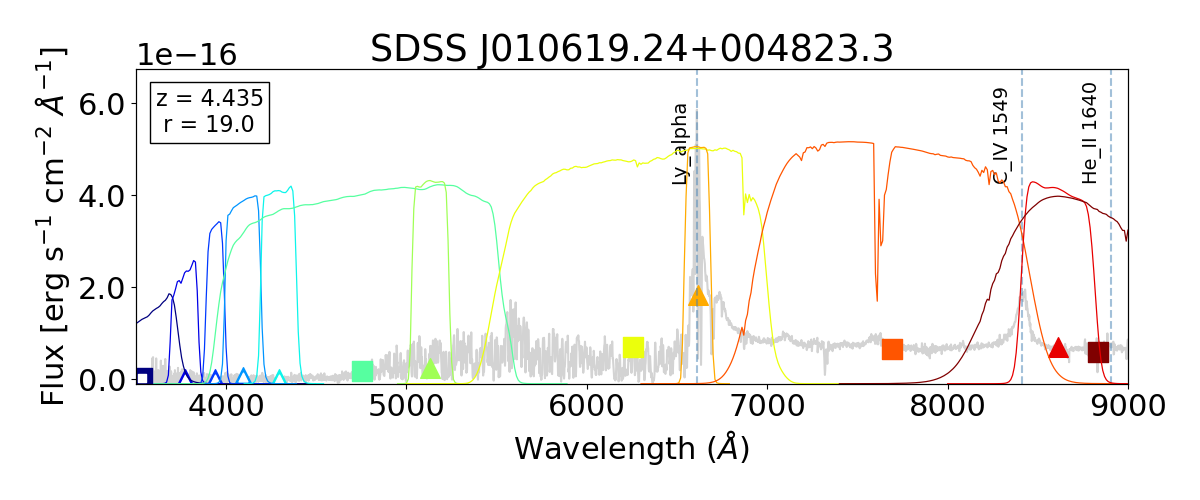}}

  \caption{Spectral energy distribution of five quasars with spectroscopic redshifts of 1.639, 2.238, 2.516, 3.207, and 4.435 (top to bottom) showing an emission line being detected at an S-PLUS narrow band. In gray, we plot the spectra observed by the SDSS programs, with the identified emission lines being plotted in vertical dashed lines. The coloured triangles (squares) represent the flux measured in the S-PLUS narrow (broad) bands, from left to right: $u$, $J0378$, $J0395$, $J0410$, $J0430$, $g$, $J051$5, $r$, $J0660$, $i$, $J0861$, $z$. The filter responses are shown accordingly. Empty triangles/squares represent the error in AB magnitude above 0.5 or non-detection at that particular band.}
  
    \label{fig:spectra_qsos}
\end{figure}
   
In Fig. \ref{fig:spectra_qsos} we show the SEDs of five examples of quasars from our sample with redshifts of 1.639, 2.238, 2.516, 3.207, and 4.435. The figure shows how the S-PLUS SEDs are shaped when one or more emission lines fall in the region of a narrow band. For instance, for $z \sim 1.63$, CIV is detected in $J0410$; for $z \sim 2.24$, Ly$\alpha$ is detected in $J0395$; for $z \sim 2.52$, Ly$\alpha$ is detected in $J0430$; for $z \sim 3.21$, Ly$\alpha$ is detected in $J0515$. At the red side of the optical wavelength range, S-PLUS has the $J0660$ narrow band with a magnitude limit of about 21.1 mag, which is deeper than all other narrow bands, and deeper than the $z$ band (see Table \ref{tab:mag}). Note from the bottom panel of Fig. \ref{fig:spectra_qsos} that the Ly$\alpha$ can be detected at the $J0660$ band in $z \sim 4.4$. Considering the large area of the southern sky that will be surveyed by S-PLUS, it will play an important role in finding quasar candidates that are under-represented in the SDSS spectroscopic sample. 

\subsection{The Wide-field Infrared Survey Explorer (WISE)}
The WISE (\citealt{2010AJ....140.1868W}) is an all-sky infrared survey that obtained photometry in four bands, in the wavelength range from 3.4 $\mu$m to 22 $\mu$m.  
In this work, we use the W1 and W2 bands from the unWISE catalogue \citep{2019ApJS..240...30S}, with effective wavelengths 3.4 $\mu$m and 4.6 $\mu$m, respectively.
The addition of these two infrared bands is known to significantly improve the performance of photo-\textit{z}s for quasars (e.g. \citealt{2012ApJ...749...41B}, \citealt{2013ApJ...772..140B}, \citealt{2015MNRAS.452.3124D}, \citealt{2017AJ....154..269Y}). 
WISE magnitudes are reported on the Vega System and their conversion to AB system is given by Eq. \ref{eq:vega}:
\begin{equation}
    m_{\text{AB}} = m_{\text{vega}} + \Delta{m},
    \label{eq:vega}
\end{equation}

where $\Delta{m} = 2.699$ and $\Delta{m} = 3.339$ for W1 and W2, respectively. It is recommended to subtract a 4 mmag and a 32 mmag offset to the unWISE measurements \citep{2019ApJS..240...30S}. The Vega to AB offsets, as constants, do not affect machine learning photo-\textit{z} methods as long as the feature space for the training set and for the unlabelled data are defined in the same way. Therefore, we do not convert these magnitudes to the AB system for this work. To apply our trained models in all sources observed in S-PLUS, we perform a CDS cross-match with the unWISE catalogue within 2\arcsec. There are 75\% out of 39\,168\,373 ($14 < r < 21.3$) sources with WISE counterpart in the S-PLUS DR4 area but only 54\% have detection in the W2 band.

\subsection{The Galaxy Evolution Explorer (GALEX)}

The GALEX (\citealt{2005ApJ...619L...1M}) is an all-sky ultraviolet survey 
that obtained photometry in two bands: far-UV ($FUV$) and near-UV ($NUV$). We retrieved data for both bands
from GALEX DR6+7 \citep{2017ApJS..230...24B} using CDS cross-match within 2\arcsec. 
The addition of these two UV bands is known to significantly improve the performance of photo-\textit{z}s for quasars (e.g. \citealt{2008ApJ...683...12B}, \citealt{2012ApJ...749...41B}, \citealt{2013ApJ...772..140B}). There are 8\% out of 39\,168\,373 ($14 < r < 21.3$) sources with GALEX counterpart in the S-PLUS DR4 area but only 2\% have measurements in the FUV band.

\subsection{Spectroscopic quasar sample}

\label{sec:spec}
We cross-matched SDSS DR16Q with S-PLUS \dr{} within 1\arcsec{} radius, using data from the SDSS Stripe 82 equatorial region for our analysis ($\sim$300 deg$^2$) resulting in 33\,151 matches. Stripe 82 region dominates the current overlap between SDSS DR16Q and S-PLUS \dr{}, and represents the best-studied region in the SDSS footprint. We selected a total of 33\,151 quasars with $r\leq22$ and $z\leq5$ for our analyses, which were cross-matched with unWISE and GALEX within 2\arcsec{}. A fraction of 93.1\% and 85.9\% out of 33\,151 have detection in $W1$ and $W2$ bands from unWISE, respectively. For GALEX, only 30.8\% and 11.1\% have detection in $NUV$ and $FUV$ bands, respectively. Note that GALEX does not provide full sky coverage and therefore some of the missing-band values are due to non-observation. Ideally, non-detection and non-observation cases should be distinguished in the training process but determining the exact fraction of non-observation is not a trivial task. For the time being, we treat all non-observation missing values as they were non-detection (see \S\ref{ssec:missing} for further details).

\raggedbottom

\section{Methods}
\label{sec:methods}
In this work, we assess the importance of narrow-band photometry in improving the quality of quasar photo-\textit{z}'s predictions. First, we discuss how much is the photo-\textit{z} prediction improved when narrow-band photometry is added into a feature space that only contains broad-band photometry. In order to answer this question, we compare single-point estimates from Random Forest experiments trained with fixed hyperparameters and data sets using two different feature colour spaces:

\begin{itemize}
\item \textbf{\texttt{broad}}: $u-r$, $g-r$, $r-i$, $r-z$
\item \textbf{\texttt{broad+narrow}}: the above four broad-band colours extended with the following seven 
narrow-band colours:
$J0378-r$, $J0395-r$, $J0410-r$, $J0430-r$, $J0515-r$, $r-J0660$, and $r-J0861$. 
\end{itemize}

We then discuss how the improvement factor changes when the GALEX and WISE broad-band photometry is also available. Similarly, as before, we compare two experiments trained with the following feature spaces:

\begin{itemize}
\item \textbf{\texttt{broad+GALEX+WISE}}: $u-r$, $g-r$, $r-i$, $r-z$, $FUV-r$, $NUV-r$, $r-W1$, $r-W2$
\item \textbf{\texttt{broad+GALEX+WISE+narrow}}: the above eight broad-band colours extended with the same seven narrow-band colours as in the first case.  
\end{itemize}

As GALEX and WISE bands improve photo-\textit{z}'s predictions even when narrow-band colours are not included (see Section \ref{sec:results}), we only consider the \texttt{broad+GALEX+WISE} and \texttt{broad+GALEX+WISE+narrow} feature spaces for further analyses. Testing three independent methods allows us to check the consistency of our conclusions about the importance of narrow-band photometry. Moreover, different methods can generally learn different patterns from the data and thus we also evaluate an ensemble of the three methods. Lastly, we also estimate photo-\textit{z}'s probability density functions (PDFs) using \flex{} and Bayesian Mixture Density Network.

\begin{figure}
\includegraphics[width=0.45\textwidth]{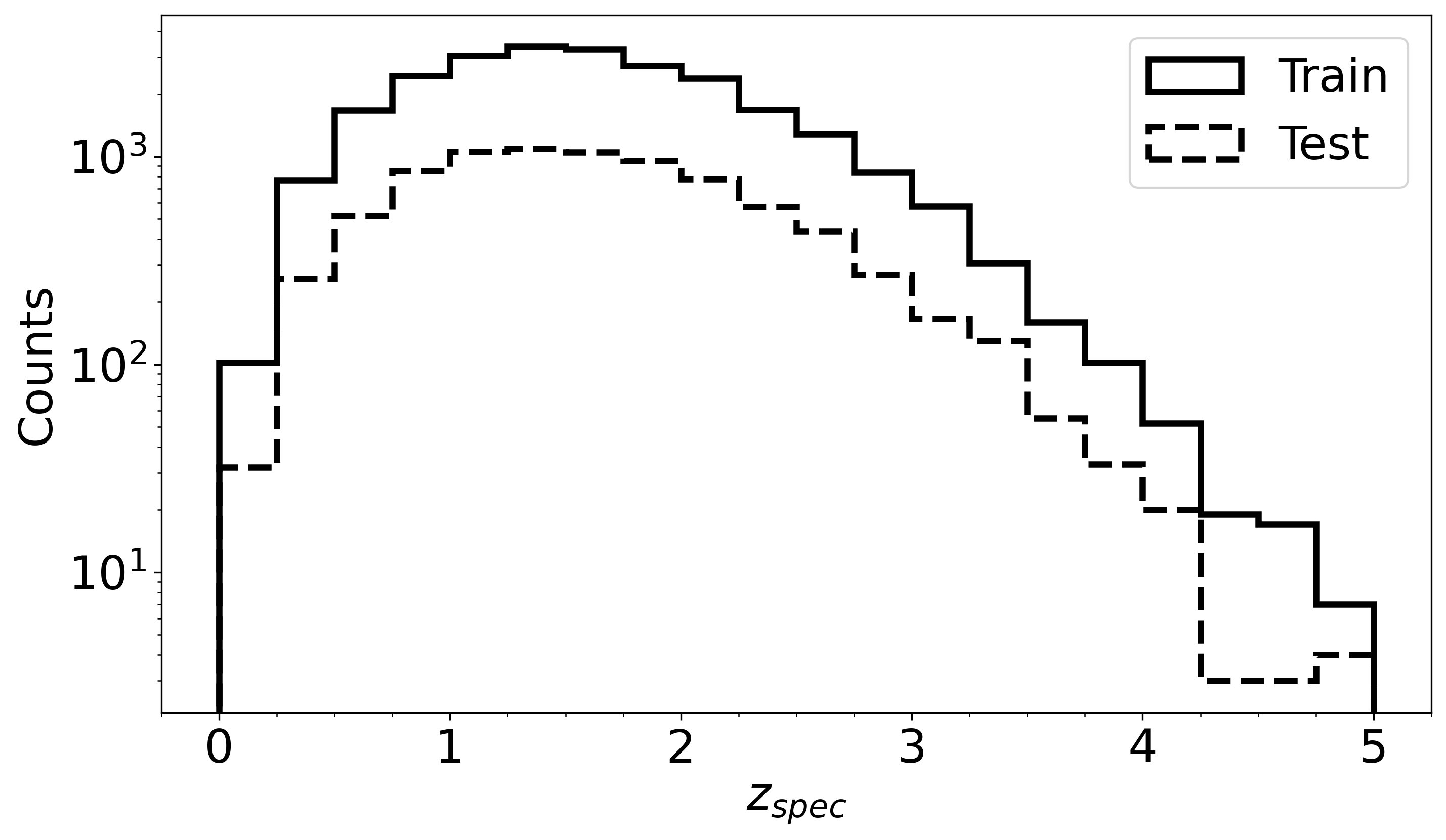}
\caption{Distributions of spectroscopic redshift for the training and testing samples.}
\label{fig:train_test_distribution}
\end{figure}

The quasar sample described in Section \ref{sec:surveys} is randomly split into training (75\%) and testing (25\%) sets. Their spectroscopic redshift distribution is shown in Fig. \ref{fig:train_test_distribution}. Note that there are only few objects in the testing set for $z>4.2$, which may affect the performance estimation in this range.  Here, we introduce the methods used in this work: Random Forest (\S\ref{ssec:rf}), \flex{} (\S\ref{ssec:flex}), and Bayesian Mixture Density Network (\S\ref{ssec:bnn}). In \S\ref{ssec:missing} we discuss how we handled missing-band values and outliers. Finally, in \S\ref{ssec:metrics} we discuss the metrics considered in this work to evaluate our model performance.

\subsection{Random Forest}
\label{ssec:rf}
Random Forest (\rf) is a non-parametric method that is robust to outliers, noise and overfitting. Moreover, it is fast-learning, and, in some sense, interpretable due to the computation of feature importance. These characteristics and its known ability to provide high accuracy in diverse fields of study make \rf\ an attractive algorithm. Since \rf\ has been extensively used in astronomy, for more details we simply refer the reader to \cite{Breiman2001}. Regarding the specific task of estimating photometric redshifts, please see: \citealt{2010ApJ...712..511C} (galaxies), \citealt{2021MNRAS.505.4847H} (galaxies), and \citealt{2021A&A...649A..81N} (quasars).

We use the \texttt{Python} implementation of \rf\ in \texttt{scikit-learn}\footnote{\url{https://scikit-learn.org/stable/modules/generated/sklearn.ensemble.RandomForestRegressor.html}} (version 1.1.3) package. We utilized the complete colour space (\texttt{broad+GALEX+WISE+narrow}) to run a grid search, resulting in the following configuration: \texttt{n\_estimators}$=$400, \texttt{min\_samples\_split}$=$2, \texttt{min\_samples\_leaf}$=$2, \texttt{max\_depth}$=$\texttt{None}, \texttt{bootstrap}$=$\texttt{True}. The photometric redshift is estimated as the mean value of the 400 predictions. We set \texttt{random\_state} to 47 for all processes, in order to enable reproducibility. The remaining hyperparameters were kept at default values of \texttt{scikit-learn} v1.3.0. In this work, we do not provide an estimated probability density function for the photometric redshift with RF.

\subsection{\flex{}}
\label{ssec:flex}
 \flex{} (\citealt{10.1214/17-EJS1302,dalmasso2020conditional}) is a non-parametric method that estimates conditional densities.
 \flex{} was applied to estimate photometric redshift distributions in the mock data produced for Vera Rubin Observatory's Legacy Survey of Space and Time (LSST; \citealt{2020MNRAS.499.1587S}), providing the best Conditional Density Estimation (CDE) loss (see Section \ref{ssec:metrics} for details)  among several other methods.  It is implemented in \texttt{R} (\citealt{rafael_izbicki_2019_3366065}, the implementation used in this work), and \texttt{Python} \citep*{taylor_pospisil_2019_3364860}.

Let $f(z|\textbf{x})$ be the probability density function of the redshift $z$ of a quasar with photometric covariates $\textbf{x}$.
 The key idea of \flex{}  is to project  $f(z|\textbf{x})$ on an orthonormal basis  $(\phi_i)_{i \in \mathbb{N}}$ (we use the Fourier basis):
\begin{equation}
\label{eq:flexcode_expansion}
    f(z|\textbf{x}) = \sum_{i \in \mathbb{N}}\beta_i (\textbf{x})\phi_i(z),
\end{equation}
where 
$\beta_i(\textbf{x})$ are the expansion coefficients. By construction, these coefficients are given by
\begin{equation}
    \beta_i(\mathbf{x}) = \langle f(\cdot|\mathbf{x}),\phi_i \rangle = \int_{\mathbb{R}} \phi_i(z)f(z|\mathbf{x})dz = \mathbb{E}[\phi_i(Z)|\mathbf{x}],
\end{equation}
and thus each coefficient $  \beta_i$ can be estimated by regressing  $\phi_i(Z)$ on $\textbf{x}$ using the spectroscopic sample. In this paper, we use Random Forests to estimate each of these coefficients.
 The estimated density is then given by $\hat{f}(z|\mathbf{x}) = \sum_{i=1}^I \hat{\beta_i}(\mathbf{x})\phi_i(z)$, where $I$ is a hyperparameter that defines the total number of expansion coefficients in the model. In practice, we choose the value of $I$ that minimizes the CDE loss function computed on a validation set (Equation \ref{eq:cde}). The single-point estimate is derived as the mode of the estimated probability density function.

\subsection{Bayesian Mixture Density Network}
\label{ssec:bnn}

We call as Bayesian Mixture Density Network (\bnn{}) the combination of two types of architectures: a Bayesian Neural Network \citep{bishop97} and a Mixture Density Network \citep{bishop94}. BMDN is similar to a Dense network, where all neurons from one layer are connected to all neurons in the previous and following layers. The strength of these connections is given by weights, represented as single values.  

\begin{figure}
\centering
     \includegraphics[page =1, width=\linewidth]{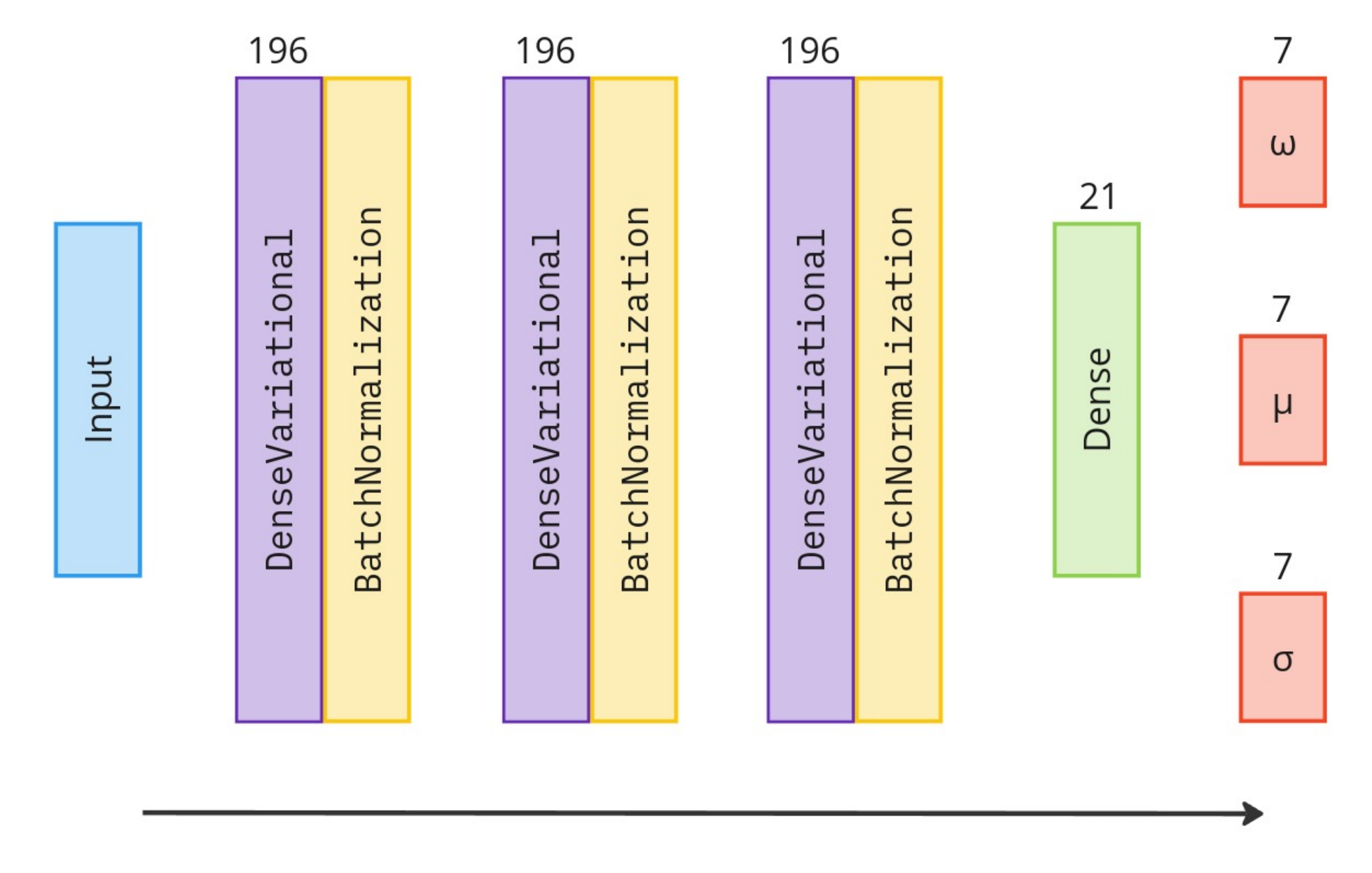}
  \caption{Architecture of the Bayesian Mixture Density Network. The input layer is followed by three blocks of \texttt{DenseVariational} and \texttt{BachNormalization} layers and a \texttt{Dense} layer. The numbers represent the number of neurons in that layer. The output layer (a \texttt{MixtureNormal}, in red) returns 7 weights ($w$), means ($\mu$) and standard deviations ($\sigma$).}
    \label{fig:bmdn_scheme}
\end{figure}

In the Bayesian framework, these single-value weights are replaced by probability distributions. In our case, we assume a Gaussian distribution. For the implementation, we used the \texttt{Tensorflow Probability} \citep{tfp}. The Mixture Density section of the network is what allows the estimation of distributions as output. This combination of architectures enables the network to represent arbitrary probability distributions while also allowing for an estimation of the epistemic (due to the model) and aleatoric (due to the data collection process) uncertainties, and provides a way to deal with the colour-redshift degeneracy (see Section \ref{ssec:pdf}). The architecture used in this work is shown in Fig. \ref{fig:bmdn_scheme}.

The \texttt{MixtureNormal} layer of the network outputs 7 Gaussian functions described by two parameters: the mean ($\mu$) and the standard deviation ($\sigma$). Each Gaussian has an associated weight ($w$), so the photo-\textit{z}'s probability density function is a combination of these Gaussian functions. For each object, we also derive a  single-point estimate by computing the mode of the estimated probability density function.

\subsection{Dealing with missing-band values}
\label{ssec:missing}
Missing-band values are intrinsic to astronomical data. We assume that any missing data in a non-detection case is "missing not at random" (MNAR; see \citealt{rubin:1976a} for more information). An example of MNAR is a high-redshift quasar that is not detected in the blue bands due to the Lyman Break (e.g. see the two bottom panels in Fig. \ref{fig:spectra_qsos}). These missing values have physical meaning and cannot be ignored. For MNAR, the removal of objects from our training set (also known as the complete-case method), or the mean/median imputation would bias our models \citep{dongpeng}.

For tree-based methods (such as Random Forest and \flex{}), we can handle the non-ignorable missing-band values by replacing them with an arbitrary out-of-the-range value (e.g. 99). The colours are calculated after the missing-band replacements. 

For the Bayesian Mixture Density Network, we first replace any missing-band values in S-PLUS by the minimum magnitude of a non-detection given in the error column for each band. Then, we standardize the data and re-scale them to the range [0,1], without taking into consideration the missing-band values in GALEX and unWISE. The last scaling process is motivated by the suggested procedure from \cite{chollet2017} of replacing missing features by zeros. Since this can only be done if the value zero is not meaningful for the problem, and since colours can have this value, we use this second scaling function to shift the distribution. 
After the scaling processes, we replace any missing-band values in GALEX and unWISE by zeros.

\subsection{Model Evaluation}
\label{ssec:metrics}

Here, we present the performance metrics considered in this work. Let $z$ be the (true) spectroscopic redshift and $\hat{z}$, the estimated photometric redshift (sometimes, especially in figures, we also refer to them as z$_{\text{spec}}$ and z$_{\text{phot}}$, respectively). 

Defining $\delta z= z - \hat{z}$,  $\delta z_{\text{norm}} = \delta z / (1+z)$ and considering $N$ being the number of evaluated objects, we define the following performance metrics for the single-point predictions:

\begin{itemize}
    \item Root-mean-squared error: \begin{equation}
    \sigma_\text{RMSE} = \sqrt{\frac{1}{N}\sum_{i=1}^{N}{(\delta z_i)^2}},
\label{eq:rmse}
\end{equation}
    \item Normalized median absolute deviation:
    \begin{equation}
\sigma_{\text{NMAD}} = 1.48 \times \text{median}\Big(\frac{|\delta z -\text{median} (\delta z)|}{1 + z}\Big)
\label{eq:nmad}
\end{equation}

\item Bias:
\begin{equation}
\mu = \overline{\delta z} = \frac{\sum_{i=1}^{N}\delta z_i}{N},
\end{equation}
\item Fraction of objects with residual above a cutoff $c$:
\begin{equation}
    \eta_{c} = \frac{N_{|\delta z_{\text{norm}}|>c}}{N}.
\end{equation}

\end{itemize}

The $\sigma_\text{NMAD}$ is commonly used to evaluate redshift estimates as it is less sensitive to outliers \citep{2019NatAs...3..212S}. Photo-\textit{z} single-point estimates are better when these metrics are closer to zero.  

Now, let $\hat{f}$ be the estimated probability density function, and let $\hat{F}$ be the estimated cumulative density function. In order to evaluate the probability density functions estimated by \flex{} and Bayesian Neural Networks, we use the following metrics:
\begin{itemize}
    \item Probability Integral Transform (PIT), which should be uniformly distributed if the density is well calibrated (\citealt*{polsterer2016uncertainphotometricredshifts}; \citealt*{freeman2017unified}; \citealt{zhao2021proc,dey2022calibrated}):
    \begin{equation}
    PIT(\hat{f},z) = 1 - \hat{F}(z \given \mathbf{x}) = 1 - \int_0^z \hat{f}(z \given \mathbf{x})dz,
    \end{equation}
\item Conditional Density Estimate (CDE) loss function \citep*{izbicki2016nonparametric,izbicki2017photo}, $\hat{L}(f,\hat{f})$, which is an estimate of:

\begin{equation}
    L(f,\hat{f}) = \int \int \left(f(z \given \textbf{x})-\hat{f}(z \given \textbf{x}) \right)^2 dz dP(\textbf{x}).
    \label{eq:cde}
\end{equation}

\end{itemize}

\raggedbottom

\raggedbottom	

\begin{table}
\caption{Performances from cross-validation with Random Forest algorithm in terms of the root-mean-squared error ($\sigma_{\text{RMSE}}$), normalized median absolute deviation ($\sigma_{\text{NMAD}}$), bias ($\mu$), and the fraction of sources with residual above 0.15 ($\eta_{0.15}$) and 0.3 ($\eta_{0.3}$). Standard deviations are typically of order 10$^{-3}$.} 
\label{tab:validation}
\centering
\setlength{\tabcolsep}{4pt}
\begin{tabularx}{\columnwidth}{@{}p{3.3cm}S[table-format=1.3]S[table-format=1.3]S[table-format=1.3]>{\raggedleft\arraybackslash}p{0.7cm}>{\raggedleft\arraybackslash}p{0.8cm}@{}}
\toprule
Colour space & {$\sigma_{\text{RMSE}}$} & {$\sigma_{\text{NMAD}}$} & {$\mu$} & {$\eta_{0.15}$} & {$\eta_{0.3}$} \\\midrule
\texttt{broad}                   & 0.647 & 0.217 &  0.001 & 0.492 & 0.225\\
\texttt{broad+narrow}            & 0.576 & 0.179 &  0.001 & 0.428 & 0.185\\
\texttt{broad+GALEX+WISE}        & 0.425 & 0.102 & -0.001 & 0.228 & 0.070\\
\texttt{broad+GALEX+WISE+narrow} & 0.410 & 0.093 &  0.001 & 0.217 & 0.066\\
\bottomrule
\end{tabularx}
\end{table}


\begin{figure}
    \includegraphics[width=0.48\textwidth]{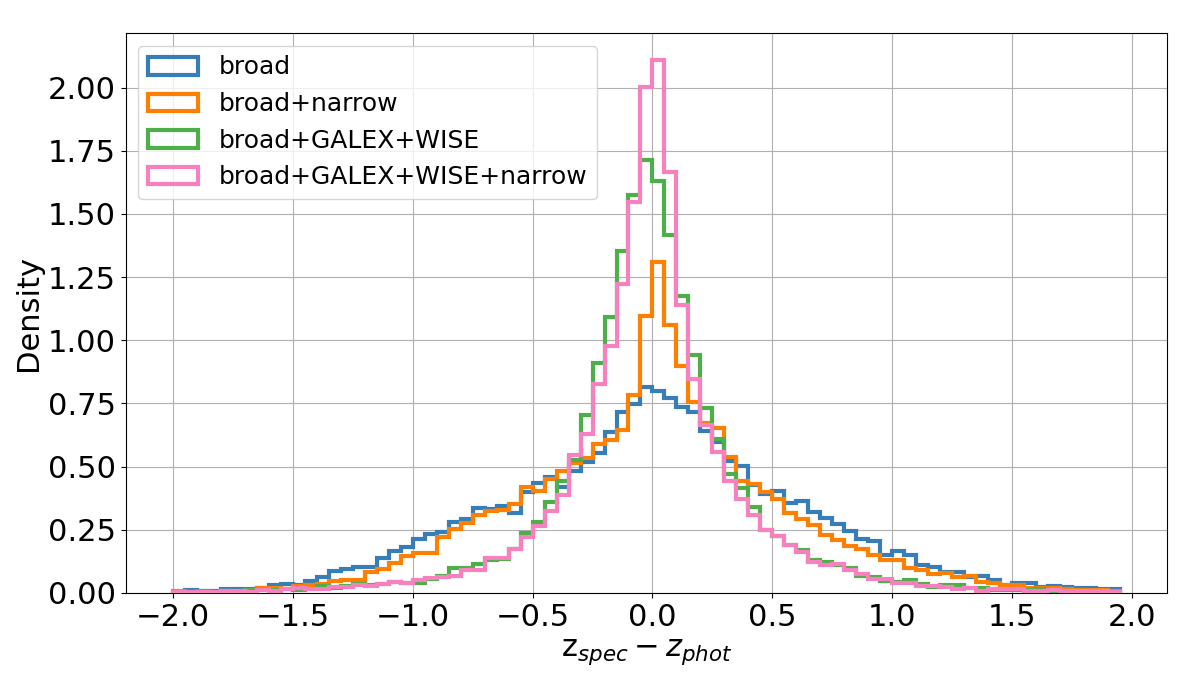}
  \caption{Distribution of the residuals ($z_{\text{spec}}-z_{\text{phot}}$) from the Random Forest model trained with \texttt{broad}, \texttt{broad+narrow}, \texttt{broad+GALEX+WISE}, \texttt{broad+GALEX+WISE+narrow}.}
    \label{fig:residuals}
\end{figure}

\section{Results}
\label{sec:results}
In this section, we analyse the importance of the narrow-band photometry for estimating quasar photometric redshifts using \rf{}, \flex{}, and \bnn{}, because different methods might learn different underlying relations between the colour space and the true redshift. We first use RF with a 5-fold cross-validation method to evaluate all feature spaces described in Section \ref{sec:methods}. After narrowing the possible choices of feature spaces, we then evaluate the single-point estimates of all the three methods (see Section \ref{ssec:single}). In Section \ref{ssec:pdf} we evaluate the estimated probability density functions from \flex{} and \bnn{}. Finally, in Section \ref{ssec:importance} we show the feature importances obtained from the tree-based models.

\subsection{Assessing the importance of narrow-band photometry}
\label{ssec:single}

 Here, we compare the performance of Random Forest experiments trained on different feature spaces with fixed hyperparameters and fixed training/validation sets.  With the choice of feature spaces described in Section \ref{sec:methods}, we aim to provide insights into the importance of narrow-band photometry for this regression problem. The quantitative comparisons were done with $\sigma_{\text{RMSE}}$, $\sigma_{\text{NMAD}}$, $\mu$, $\eta_{0.15}$, and $\eta_{0.3}$ under a 5-fold cross-validation scheme. These metrics are shown in Table \ref{tab:validation}.

When narrow-band colours are added to the broad-band colour space,  there is a decrease in the values of $\sigma_{\text{RMSE}}$ and $\sigma_{\text{NMAD}}$ by about 11\% and 17\%, respectively. As is discernible from Table \ref{tab:validation}, in all cases $\sigma_{\text{RMSE}}$ is much larger than $\sigma_{\text{NMAD}}$ due to 
the influence of outliers. The impact of outliers is also well-tracked through $\eta$ metrics. 
The $\eta_{0.15}$ metric decreased from 49\% to 43\%, while $\eta_{0.3}$ also decreased from 23\% to 19\%. The biases are sufficiently small for all cases with an order of $10^{-3}$.

\begin{table}
\setlength{\tabcolsep}{12pt}
\caption{This table shows the mean performance improvement in per cent due to the addition of narrow-band information for the subset of sources without GALEX ($FUV$ and $NUV$) or WISE ($W1$ and $W2$) counterparts.  Here we only compare \texttt{broad+narrow} and \texttt{broad} colour spaces. The improvements are slightly below the 11\% ($\sigma_\text{RMSE}$) and 17\% ($\sigma_\text{NMAD}$) calculated from Table \ref{tab:validation}. For completeness, we also show the results for the complementary subsets. }
\label{tab:val_subset}
\begin{tabular}{@{}lccc@{}}
\toprule
Sample Subset  & \begin{tabular}[c]{@{}l@{}} Decrease in \\ $\sigma_{\text{RMSE}}(\%)$ \end{tabular}& \begin{tabular}[c]{@{}l@{}} Decrease in \\ $\sigma_{\text{NMAD}}(\%)$ \end{tabular} & Sample size\\\midrule
Without GALEX                    & 9.6 & 14.1 & 3405$\sim$3478\\
With GALEX                  & 15.6 & 25.4 &  1494$\sim$1567\\
Without WISE                  & 9.2 &  16.5 & 288$\sim$301\\
With WISE                   & 11.2 &  17.6 & 4642$\sim$4684\\

\bottomrule
\end{tabular}
\end{table}

\begin{figure*}
\centering
\subfloat[]{\includegraphics[width=0.45\textwidth]{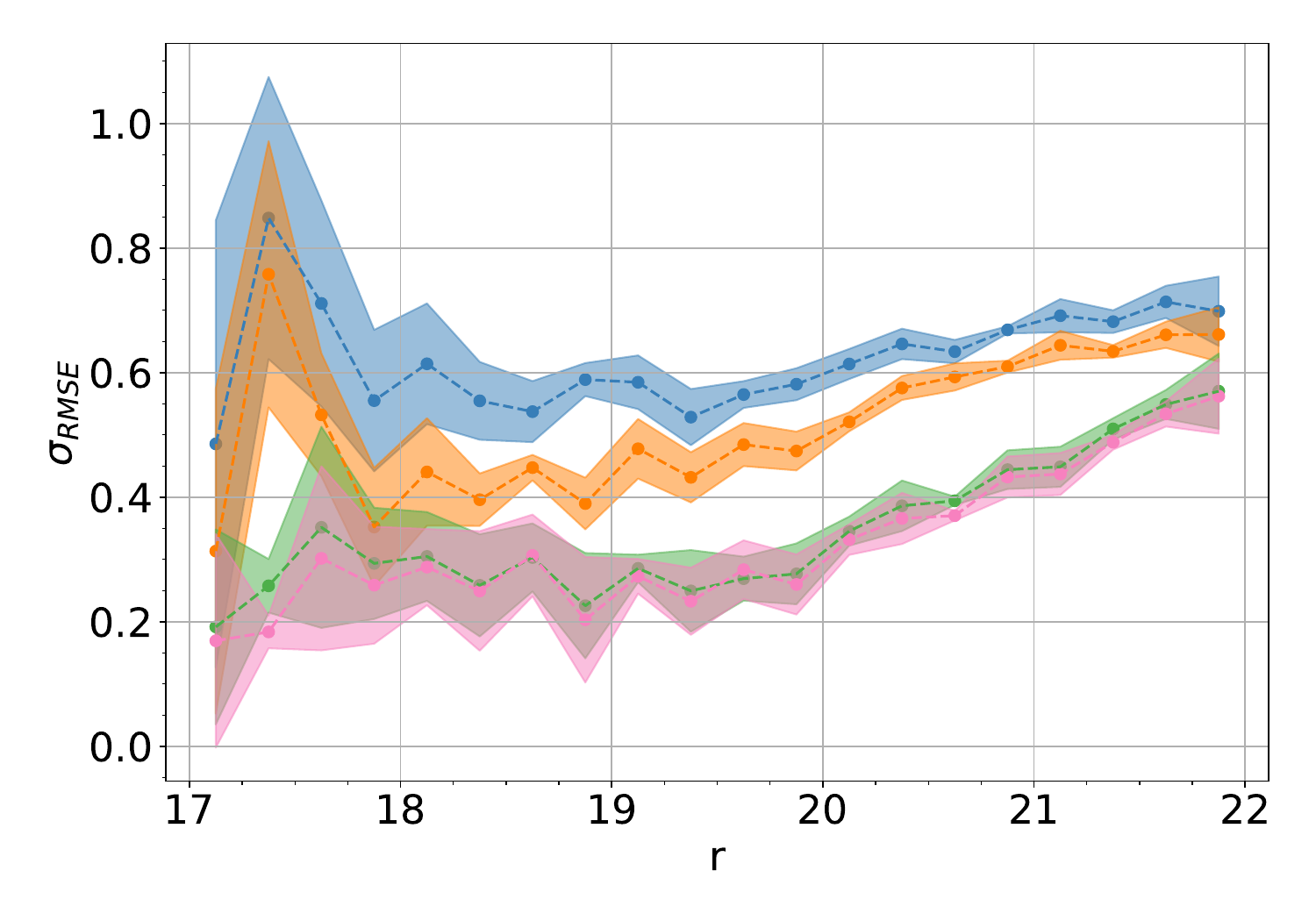}
\label{fig:metrics_a}}
\centering
\subfloat[]{\includegraphics[width=0.45\textwidth]{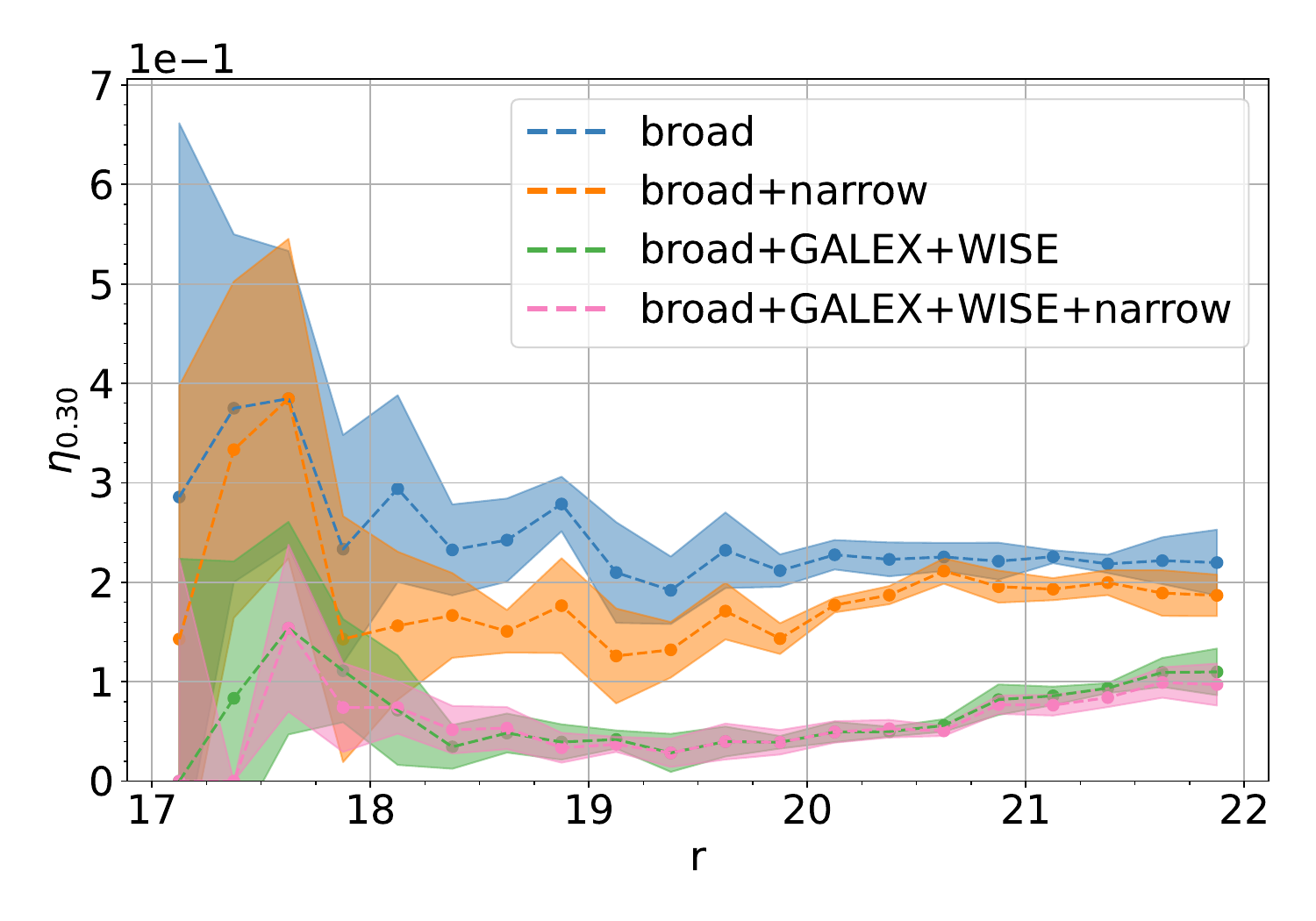}%
\label{fig:metrics_b}}

\caption{Median performance ($\pm$ 1$\sigma$) of Random Forest experiments trained on: \texttt{broad}, \texttt{broad+narrow}, \texttt{broad+GALEX+WISE}, and \texttt{broad+GALEX+WISE+narrow}.  (a) $\sigma_{\text{RMSE}}$ per magnitude $r$ and (b) $\eta_{0.30}$ per magnitude $r$.}
\vspace{-10 pt}  
\label{fig:metrics}
\end{figure*}

These improvements are smaller than the improvements observed when 
GALEX and WISE colours are added to broad-band colours (i.e. without using narrow-band colours). 
As shown in Table \ref{tab:validation}, the extended wavelength range due to the addition of
GALEX and WISE colours improve $\sigma_{\text{NMAD}}$ and $\eta_{0.15}$ metrics by as much as about a factor of two. When all the data are used, broad-band colours, GALEX and WISE colours and narrow-band colours,
only slight additional improvements are observed. These improvements are visualized using 
the histograms of the residuals ($z_{\text{spec}} - z_{\text{phot}}$) in Fig. \ref{fig:residuals}; the residual distribution is more concentrated around zero when information from all bands is included in our model. 
Fig. \ref{fig:metrics} shows how $\sigma_{\text{RMSE}}$ and $\eta_{0.30}$ vary with the apparent magnitude $r$. The addition of narrow-band and/or GALEX and WISE colours improves the regression performance, especially for brighter objects.

As discussed above, the addition of GALEX and WISE colours to broad-band colours improves
the performance of photo-\textit{z} more than the addition of narrow-band colours to broad-band colours. 
However, we emphasize that GALEX and WISE bands are effectively shallower than the S-PLUS bands
and for faint objects, only narrow-band photometry is available, besides broad-band photometry. To address these cases where narrow-band photometry will play a more important role, we calculate the metrics on subsets of the validation folds that do not have GALEX or WISE counterparts (see Table \ref{tab:val_subset}).  For $\sigma_\text{RMSE}$ the improvement is of an order of $\sim$9\% for both cases. Regarding $\sigma_\text{NMAD}$, the improvement is 14.1\% for sources without GALEX counterpart, and 16.5\% for sources without WISE counterpart.

For further analyses that include \flex{} and BMDN, we only consider the \texttt{broad+GALEX+WISE} and \texttt{broad+GALEX+WISE+narrow} colour spaces. In Table \ref{tab:final} we show the metrics for all three methods; they all show small improvements from the addition of narrow bands regarding $\sigma_\text{RMSE}$, $\eta_{0.15}$, and $\eta_{0.30}$.
There is an interesting behaviour observed for $\sigma_{\text{NMAD}}$: without narrow bands, all three methods have similar performance but when narrow bands are added, \flex{} and BMDN outperform RF method by a factor of two, approximately.
On the other hand, RF method achieves about an order of magnitude smaller bias in comparison to FlexCoDE and BMDN for all models, except for the BMDN with the narrow bands.
When we average the predictions from these three models, $\sigma_\text{RMSE}$ and $\eta_{0.30}$ are improved compared to each individual model.
Note that the first two rows in Table \ref{tab:final} are not identical to the last 
two rows in Table \ref{tab:validation} because the former is based on the testing set 
while the latter is based on the validation sets.

\begin{table*}
\setlength{\tabcolsep}{16.7pt}
\caption{Final performances calculated on the test set for Random Forest, \flex{} and Bayesian Mixture Density Network. The last two rows correspond to the performance when we average the single-point estimates from these three models. The best metric values are highlighted in bold. }
\begin{tabular}{@{}lllllll@{}}
\toprule
& Feature Space & \multicolumn{1}{c}{$\sigma_{\text{RMSE}}$} & \multicolumn{1}{c}{$\sigma_{\text{NMAD}}$} & \multicolumn{1}{c}{$\mu$} & \multicolumn{1}{c}{$\eta_{0.15}$} & \multicolumn{1}{c}{$\eta_{0.30}$} \\\midrule
\multirow{2}{*}{Random Forest}                    & Without narrow bands & \RFrmse     & \RFnmad      & \RFbias     & \RFnquinze     & \RFntrinta     \\
                                                  & With narrow bands    & \RFnbrmse   &  \RFnbnmad   & \RFnbbias   & \RFnbnquinze   & \RFnbntrinta   \\\midrule
\multirow{2}{*}{FlexCoDE}                         & Without narrow bands & \FLEXrmse   & \FLEXnmad    & \FLEXbias   & \FLEXnquinze   & \FLEXntrinta   \\
                                                  & With narrow bands    & \FLEXnbrmse &  \FLEXnbnmad & \FLEXnbbias & \FLEXnbnquinze & \FLEXnbntrinta \\\midrule
\multirow{2}{*}{Bayesian Mixture Density Network} & Without narrow bands & \BNNrmse    & \BNNnmad     & \BNNbias    & \BNNnquinze    & \BNNntrinta    \\
                                                  & With narrow bands    & \BNNnbrmse  & \BNNnbnmad   & \BNNnbbias  & \BNNnbnquinze  & \BNNnbntrinta  \\\midrule
\multirow{2}{*}{Average}                          & Without narrow bands & \AVErmse    & \AVEnmad     & \AVEbias    & \AVEnquinze    & \AVEntrinta    \\
                                                  & With narrow bands    & \AVEnbrmse  &  \AVEnbnmad  & \AVEnbbias  & \AVEnbnquinze  & \AVEnbntrinta  \\\bottomrule
\end{tabular}
\label{tab:final}
\end{table*}

\subsection{Probability density functions}
\label{ssec:pdf}
\begin{figure*}
 \centering
 \begin{tabular}{c}
    \includegraphics[trim={0 0 0 0},clip, width=0.96\textwidth ]{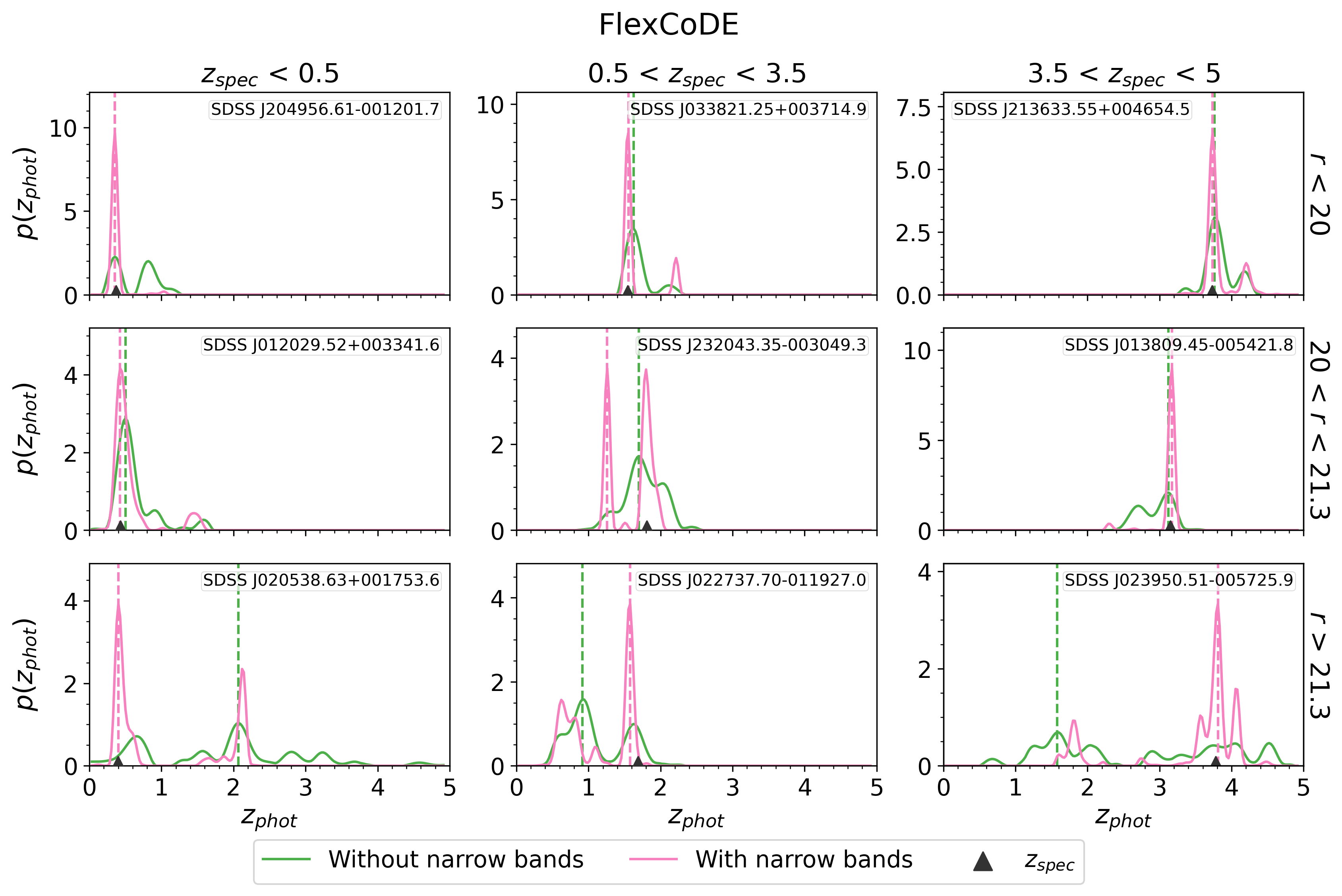} \\

        \includegraphics[trim={0 1.3cm 0cm 0},clip, width=0.96\textwidth ]{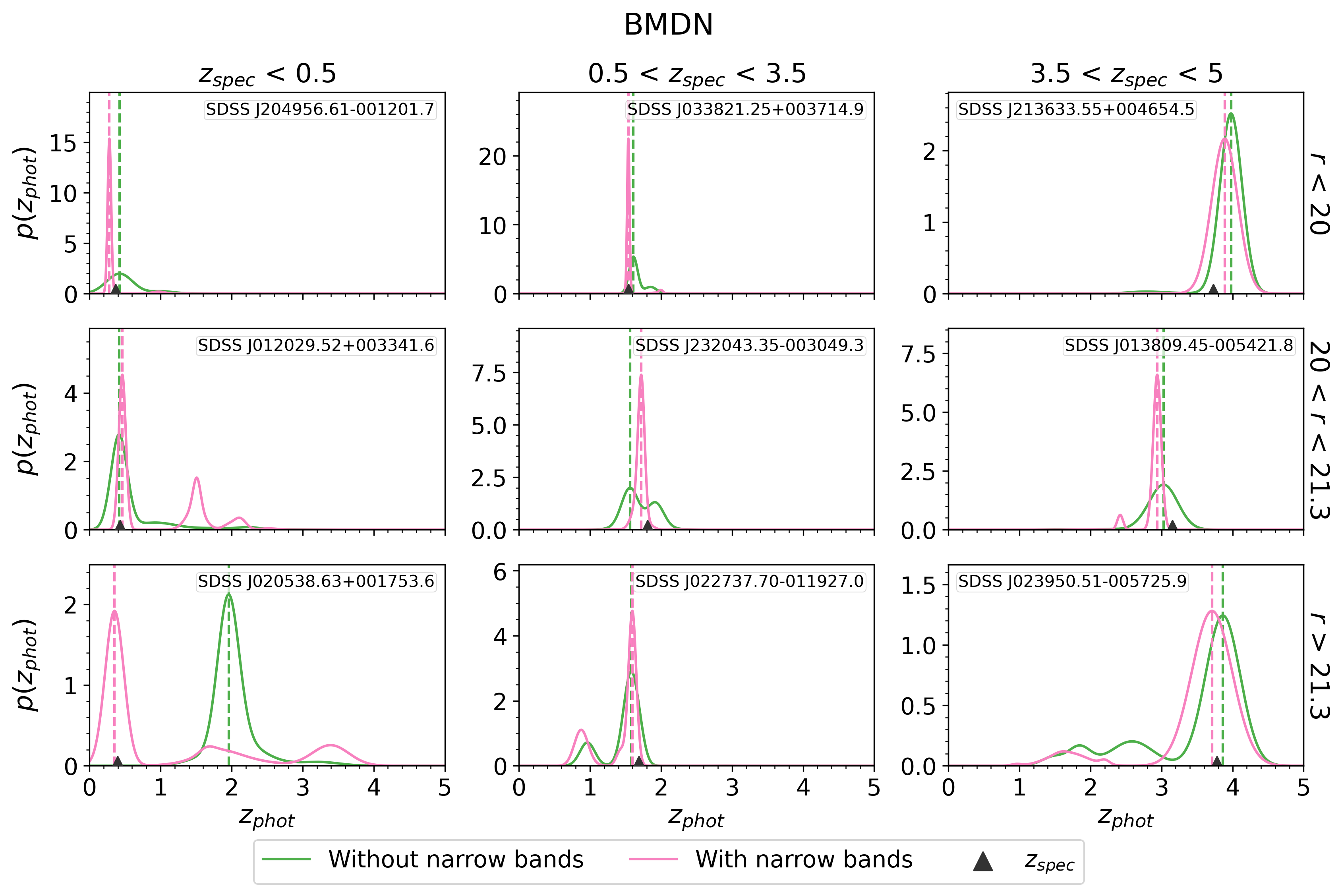}\\
    \end{tabular}
  \caption{Density estimations $p(z)$ and single-point estimates from \flex{} and \bnn{} trained with \texttt{broad+GALEX+WISE} (green) and \texttt{broad+GALEX+WISE+narrow} (pink) for the photometric redshift of 9 quasars sampled from the testing set. The spectroscopic redshift is pointed out with a triangle marker. The single-point estimates are shown with dashed vertical lines.} 
    \label{fig:flexcode_pdf}
\end{figure*}

Performance analyses from single-point estimates, however, do not account for the uncertainties in the photo-\textit{z} predictions. It is expected that the photo-\textit{z}'s PDFs can be multimodal due to the colour-redshift degeneracy, which can be seen in Figure \ref{fig:flexcode_pdf} for a random selection of nine quasars from the test set at different ranges of spectroscopic redshift and magnitude. For instance, the colours of the quasar SDSS J232043.35-003049.3 are related to, approximately, two apart single-point estimates for \flex{} narrow-band model (i.e. \texttt{broad+GALEX+WISE+narrow}), for which the PDF's second highest peak is the one centred on the correct value. Nevertheless, the PDFs' primary peak from \flex{} with the narrow-band model covers well the ``true'' redshift for most of the examples shown.

In comparison with \bnn{}, the \flex{} PDFs are generally more accurate for these particular model and examples. It is notable, however, that the PDFs are less certain on the predicted photo-\textit{z} (i.e. distributions present more peaks) for the cases in the tail of the spectroscopic redshift distribution of the training sample beyond the $r$ magnitude depth of S-PLUS ($r>21.3$).  This behaviour is also seen with the broad-band model (i.e. \texttt{broad+GALEX+WISE}) and therefore might be related to the increasing photometric uncertainties at fainter magnitudes. Although the PDFs show more confusion at this range, the single-point prediction still agrees fairly well to the ``true'' redshift but only with the narrow-band model.

\begin{figure}
    \subfloat[]{\includegraphics[width=0.48\textwidth]{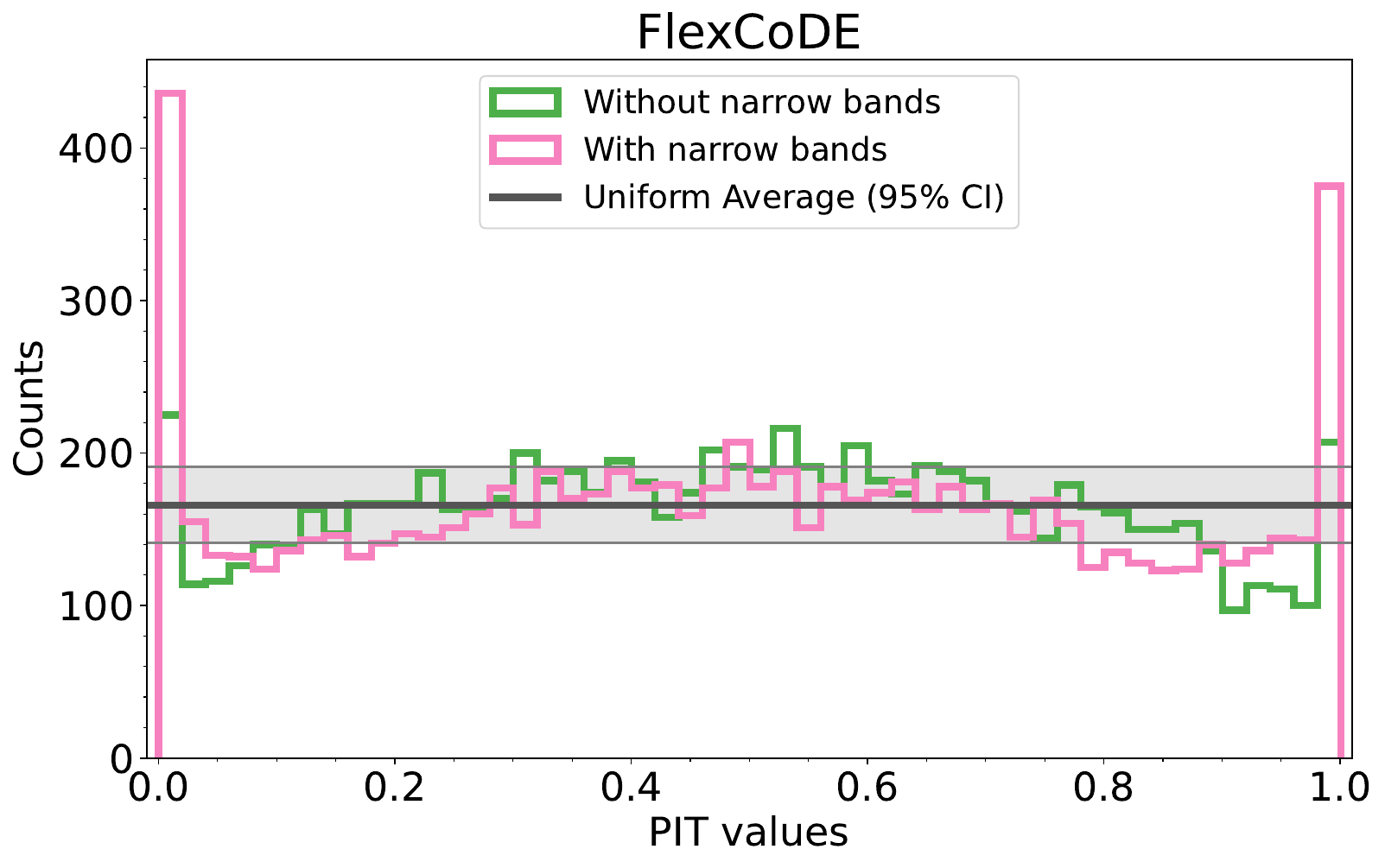}}
    
    \subfloat[]{\includegraphics[width=0.48\textwidth]{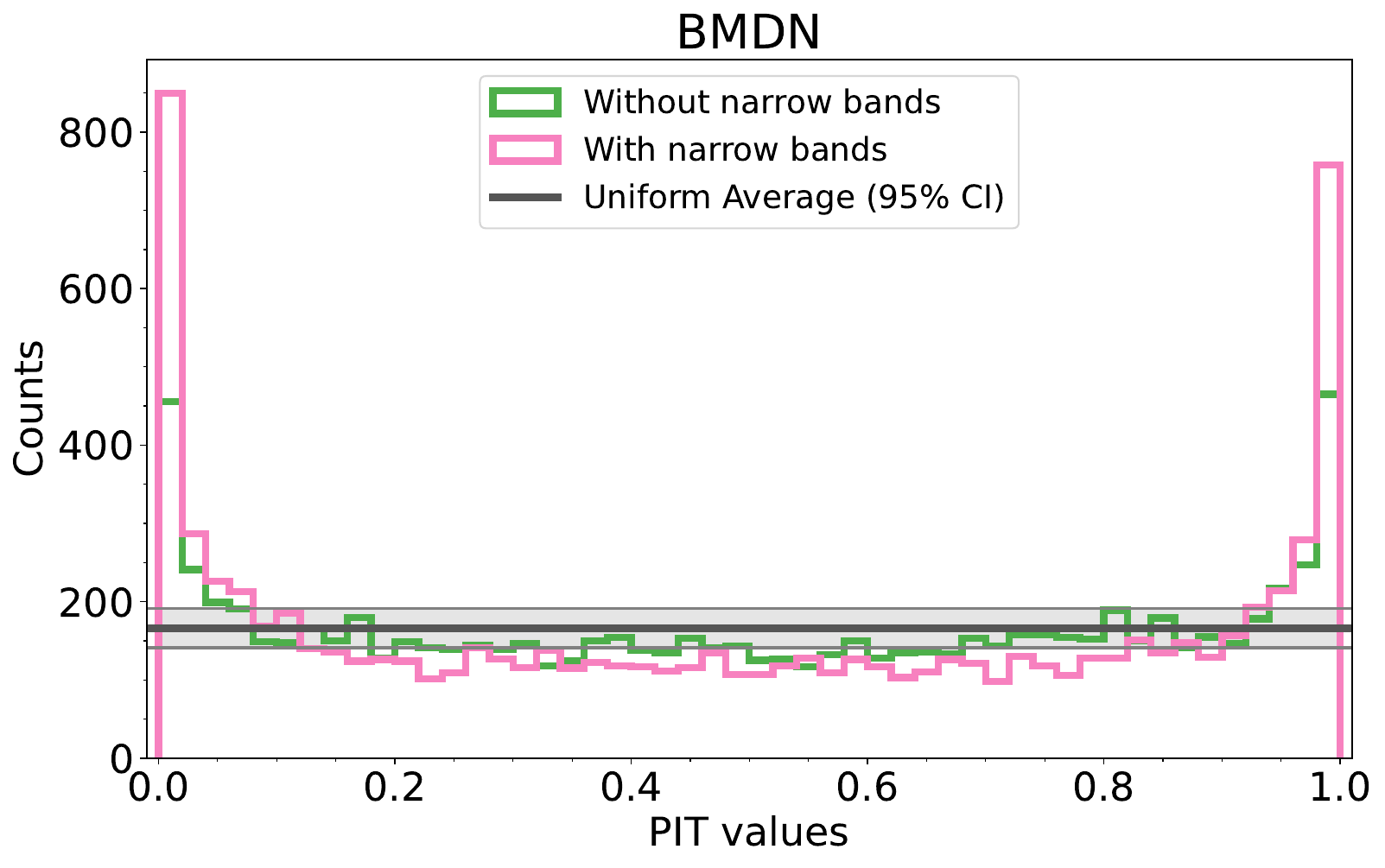}}
    
  \caption{PIT distribution for (a) \flex{} trained without narrow bands (\texttt{broad+GALEX+WISE}) and with narrow bands (\texttt{broad+GALEX+WISE+narrow}); (b) \bnn{} trained without narrow bands and with narrow bands. The PDFs are considered well-calibrated if the distribution of PIT is described by a Uniform distribution. }
    \label{fig:pit}
\end{figure}

Other quantitative comparisons should also be considered in order to evaluate the whole PDF information, such as the PIT distribution and the CDE loss function. As we have seen in Fig. \ref{fig:flexcode_pdf}, narrow-band model PDFs are generally sharper than broad-band model PDFs for both \flex{} and \bnn{}, suggesting that the predicted distribution is more precise (but not necessarily more accurate) when narrow-band information is included in the model. On the other hand, this could mean that the predictive uncertainties are not being properly accounted for and the distributions are sharper than they should be. The high edges (U-shape) of the PIT distributions shown in Fig. \ref{fig:pit} agree with the suggestion of underdispersed PDFs. From this figure, we see that the narrow-band PDFs are more underdispersed than the broad-band PDFs. It then requires a better calibration of these PDFs that can either be achieved with a different machine learning architecture or with a recalibration process \citep[e.g.][]{2021arXiv211015209D}, which is planned to be pursued in future work.  

Nevertheless, we showed good evidence that the inclusion of narrow-band photometry is valuable for the precision of photometric redshift estimates. This is also supported by the calculated CDE losses, for which each individual PDF is evaluated (note that the loss can be estimated even if the true PDF is unknown). The CDE loss functions for \flex{} are $-3.27$ and $-1.36$ when trained with and without narrow bands, respectively. For \bnn{}, these values are  $-2.47$ and $-1.31$. Therefore, the \flex{} narrow-band model provided the best photo-\textit{z} predictions during the testing process.

Given that the feature space and the training sample are the same, the differences in PDF shapes are model intrinsic. Different algorithms can learn different underlying patterns from the data, hence these PDFs could be combined to provide more powerful information. Assessing the possible combination of these PDFs is beyond the scope of this paper, but we provide the PDFs from both methods in our value-added catalogue (see Section \ref{sec:vac}). Based on the CDE loss, the \flex{} trained with narrow bands delivered the best photo-\textit{z}s and is our general recommendation for broad usage.

\subsection{Feature importances from tree-based models}
\label{ssec:importance}

\begin{figure}
\centering
    \subfloat[]{
    \includegraphics[width=0.2\textwidth]{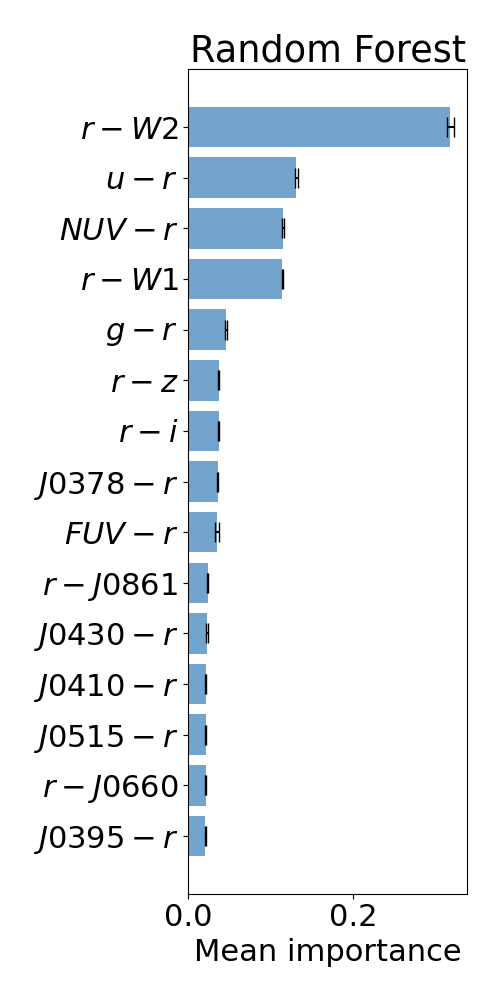}}
    \subfloat[]{
    \includegraphics[width=0.2\textwidth]{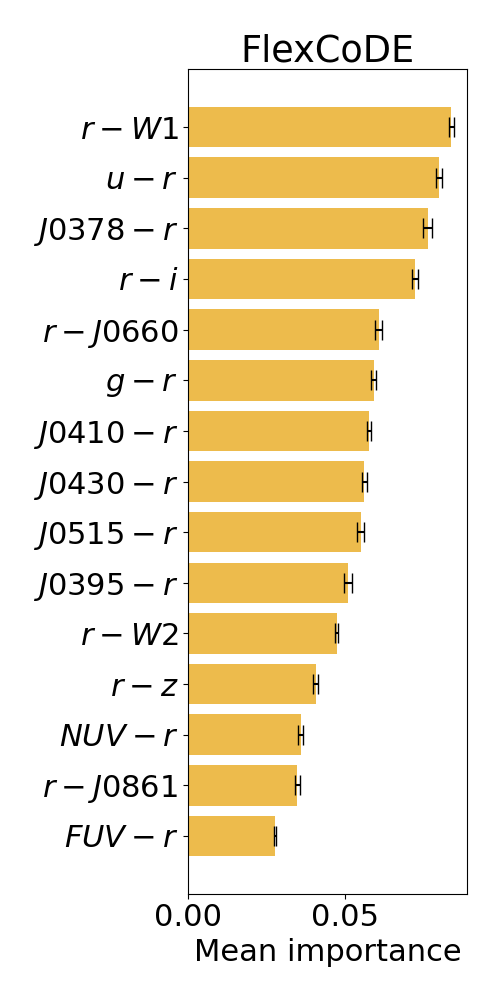}}

  \caption{Average estimated importances and standard deviation for (a) Random Forest and (b) FlexCoDE.}
    \label{fig:importances}
\end{figure}

In Fig. \ref{fig:importances}, we show the estimated feature importances for the Random Forest and the \flex{} models. In both models, the top two most important features for the quasar photo-\textit{z} estimation are the same or strongly correlated. In first place, we have $r-W1$ for \rf{}, and $r-W2$ for \flex{}. These two features are strongly correlated with Pearson correlation of 0.98. In the second place, the colour $u-r$ appears for both methods. These features ($r-W1$, $r-W2$, and $u-r$) measure the overall shape of the SED over the wide wavelength range. 
Beyond the second place, there is no interpretation agreement between the two models. While the first narrow-band ($J0378-r$) appears in 8th place for Random Forest, this same colour is the third most influential feature for \flex{}. This difference could be explained by the correlations between some of the colours, which can mislead the interpretation of the feature importances and does not necessarily mean that less important features are not relevant. Interestingly, not only do narrow-band colours have higher importances for \flex{} but also the overall importances are more equally distributed than \rf{}. The importances shown in this figure might also be closely related to the relative depths of each band (see Table \ref{tab:mag}) but further investigation would be needed. Note that the interpretation of these importances is limited to understanding what features had the most influence in these particular prediction models and no physical or general conclusions can be taken from those. Nevertheless, there are good indications of the narrow-band importance in predicting more accurate photo-\textit{z}s, as suggested by the \flex{} importances estimates and, especially, considering the CDE loss estimates discussed in the last section. 

\raggedbottom

\section{QUCATS: The Quasar Catalogue for S-PLUS} 
\label{sec:vac}

\begin{table*}
\setlength{\tabcolsep}{36pt}
\caption{Description of the information provided in QuCatS. All columns here described are type \texttt{float32}. We also include useful information from the main survey (\texttt{ID}, \texttt{Field}, \texttt{RA}, \texttt{DEC}, PStotal magnitudes and their corresponding errors) and the classification value-added catalogues (\texttt{PROB\_QSO}, \texttt{PROB\_STAR}, \texttt{PROB\_GAL}, \texttt{model\_flag)}. These extra columns are described in \url{https://splus.cloud/documentation/DR4}.}
\label{tab:vac}
\begin{tabular}{@{}lll@{}} 
\toprule
Column name & Description \\ \midrule
z\_rf      &  Photo-\textit{z} estimated with \rf{}         \\
z\_bmdn\_peak      &  Photo-\textit{z} estimated with \bnn{} (peak of the PDF)           \\ 
z\_flex\_peak      &   Photo-\textit{z} estimated with \flex{} (peak of the PDF)          \\
z\_mean        &    Average of \{z\_rf, z\_bmdn\_peak, and z\_flex\_peak\}         \\ 
z\_std      &   Standard deviation of \{z\_rf, z\_bmdn\_peak, and z\_flex\_peak\}         \\
n\_peaks\_bmdn        &   Number of peaks for \bnn{}'s PDF          \\ 
z\_bmdn\_pdf\_weight\_[1-7]     & Weight of the [1-7]-th Gaussian distribution estimated with \bnn{}          \\
z\_bmdn\_pdf\_mean\_[1-7]       &  Mean of the [1-7]-th Gaussian distribution estimated with \bnn{}          \\ 
z\_bmdn\_pdf\_std\_[1-7]       &  Standard deviation of the [1-7]-th Gaussian distribution estimated with \bnn{}          \\ 
z\_flex\_pdf\_[1-200]       &  Probability for the [1-200]-th redshift within the interval [0.034, 4.913] estimated with \flex{}          \\ 
\bottomrule
\end{tabular}
\end{table*}

The Quasar Catalogue for S-PLUS (QuCatS) provided with this paper contains 645\,980 quasar candidates with classification probabilities above 80\% up to the photometric depth of $r$ band ($r<21.3$) and good photometry quality in the detection image (\hbox{SEX\_FLAGS\_DET $=0$}) that comes from \texttt{SExtractor}. For the selection of quasar candidates, we use the star/quasar/galaxy classification probabilities from \citealt{2021MNRAS.507.5847N} that were obtained with a Random Forest fit on S-PLUS and WISE magnitudes, and morphological parameters extracted from the S-PLUS images.
Sources observed in the CCD borders that were observed in multiple fields were removed from this catalogue via an internal cross-match in RA and Dec within 1\arcsec. Our catalogue covers a total of 1414 fields from S-PLUS DR4, totalling approximately 3\,000 $\text{deg}^2$ of the southern sky. This value-added catalogue is downloadable at \url{https://splus.cloud/files/QuCatS_Nakazono_and_Valenca_2024.csv} and its columns are described in Table \ref{tab:vac}. We do not make any cuts in photometric errors for the catalogue we provide in this paper but we advise users to make proper cuts for their specific science cases.

Below, we show an example of an ADQL query\footnote{Query can be executed directly at the \url{https://splus.cloud} platform, via Python using the \texttt{splusdata} package, or using TAP service with the following address \url{https://splus.cloud/public-TAP/tap}}. The selection criteria can be easily modified in this query to retrieve more (or less) sources, where \texttt{[columns]\footnote{List of columns can be found at \url{https://splus.cloud/catalogtools/tap}.}} are defined by the user. We strongly recommend that a ``\texttt{WHERE det.Field = [field]\footnote{List of all unique field names and their central position can be obtained through \url{https://splus.cloud/files/documentation/iDR4/tabelas/iDR4_pointings.csv}}}'' condition is added to this query in order to run one query per field, as big queries can overload the server.
Note that one can relax the selection criteria and retrieve more quasar candidates from the S-PLUS data base, as we derived photo-\textit{z} estimations for all objects in each field regardless of their classification. Likewise, one can be more conservative with the selection criteria in order to improve the purity of the selected sample.

\begin{verbatim}
SELECT [columns] from dr4_vacs.dr4_qso_photoz as z
JOIN dr4_vacs.dr4_star_galaxy_quasar  
as sqg on sqg.id = z.id
JOIN dr4_dual.dr4_dual_r 
as r on r.id = z.id
JOIN dr4_dual.dr4_detection
as det on det.id = z.id
WHERE sqg.PROB_QSO >= 0.8 and
r.r_PStotal < 21.3 and
det.SEX_FLAGS_DET = 0
\end{verbatim}

\begin{figure*}
\includegraphics[width=1\textwidth]{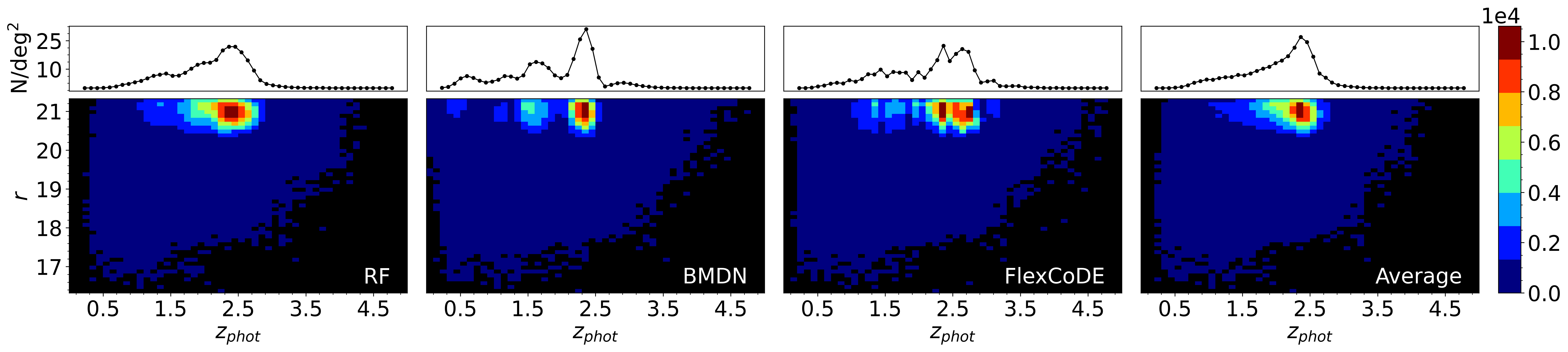}
  \caption{Density maps of photometrically selected quasar in S-PLUS DR4 with classification probabilities above 80\% \citep{2021MNRAS.507.5847N} and photometric redshift estimated with \rf{}, \bnn{}, \flex{}, and the average of these three methods. Only sources with good photometry flag ($\texttt{SEX\_FLAGS\_DET} =0$) and $r<21.3$ are plotted. The colour map indicates the number of quasar candidates in bins of 0.1 for both apparent magnitude $r$ and photo-\textit{z}. In the upper panels, we show the number of quasar candidates per squared degree for 0.1 bins of photo-\textit{z}. } 
\label{fig:density_map}
\end{figure*}

\begin{figure*}
\includegraphics[page =1, width=\linewidth]{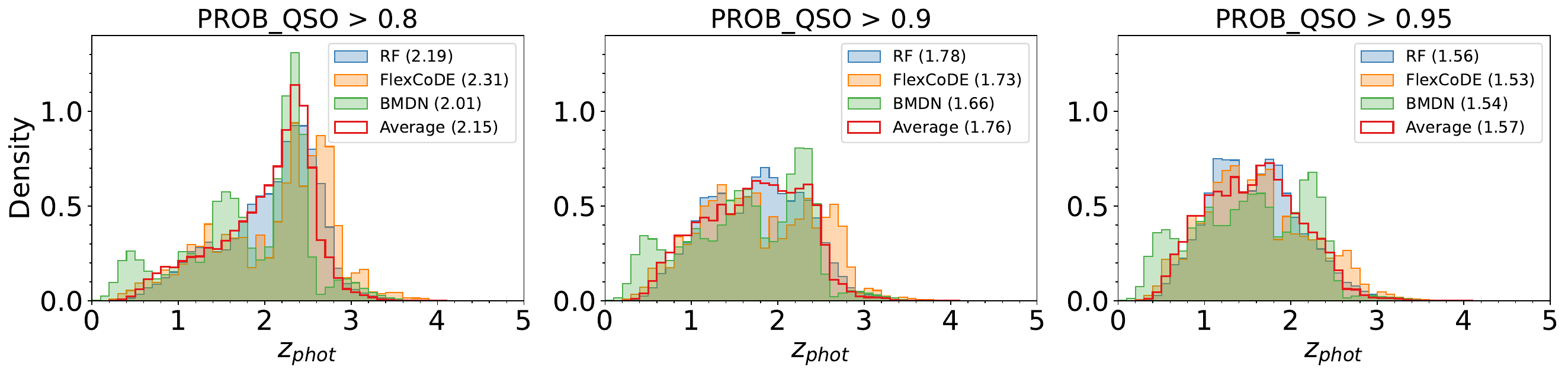}
  \caption{Distribution of photometric redshift for the sources that are photometrically classified as quasars in S-PLUS with a probability above 0.8, 0.9, and 0.95. Only sources with good photometry flag ($\texttt{SEX\_FLAGS\_DET} =0$) and $r<21.3$ are plotted. We can note that different estimation methods are leading to more alike photo-\textit{z} distributions for high-confident quasar candidates ($\texttt{PROB\_QSO} > 0.95$). The median of each distribution is given in the legend.}
    \label{fig:vac_z}
\end{figure*} 

Density maps for the selected quasar candidates show a multimodal distribution of photo-\textit{z}s for \bnn{} and \flex{}, while \rf{} deliver a smoother single-peak distribution (Fig. \ref{fig:density_map}).
In Fig. \ref{fig:vac_z} we show the photo-\textit{z} distribution for the quasar candidates in the catalogue provided in this paper with probability of being a quasar down to 0.8, 0.9, and 0.95. We can see that as we restrict our sample to more high-confident quasars, there is an increasing convergence among the photo-\textit{z} distributions estimated by the different methods considered in this work. For high-confident quasar candidates (\texttt{PROB\_QSO} $>$ 0.95), the distributions have medians at $z_\text{phot}\sim 1.55$. 

\begin{figure}
    \subfloat[]
{\includegraphics[width=0.45\textwidth]{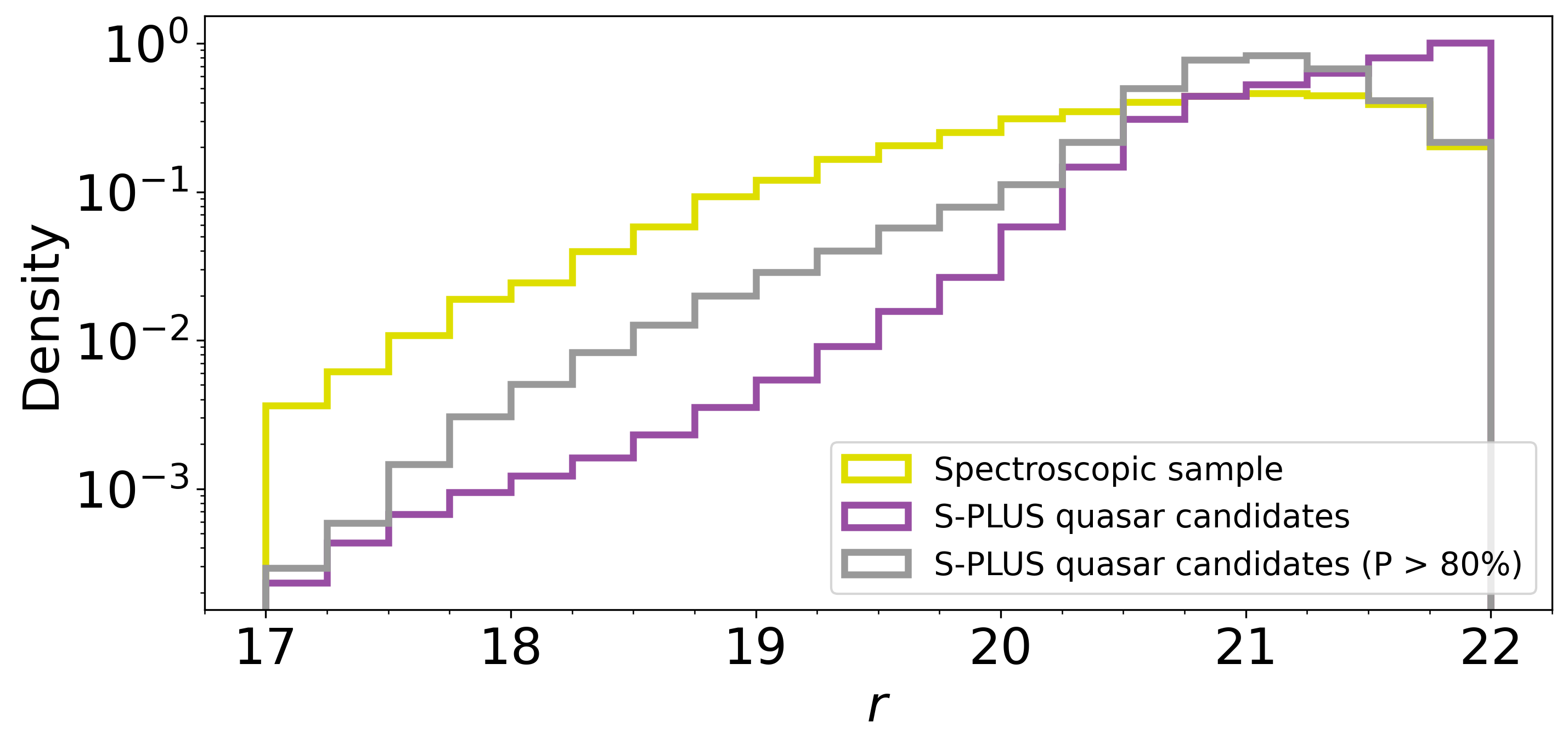}}
    \label{fig:train_test_a}
    \subfloat[]
{\includegraphics[width=0.45\textwidth]{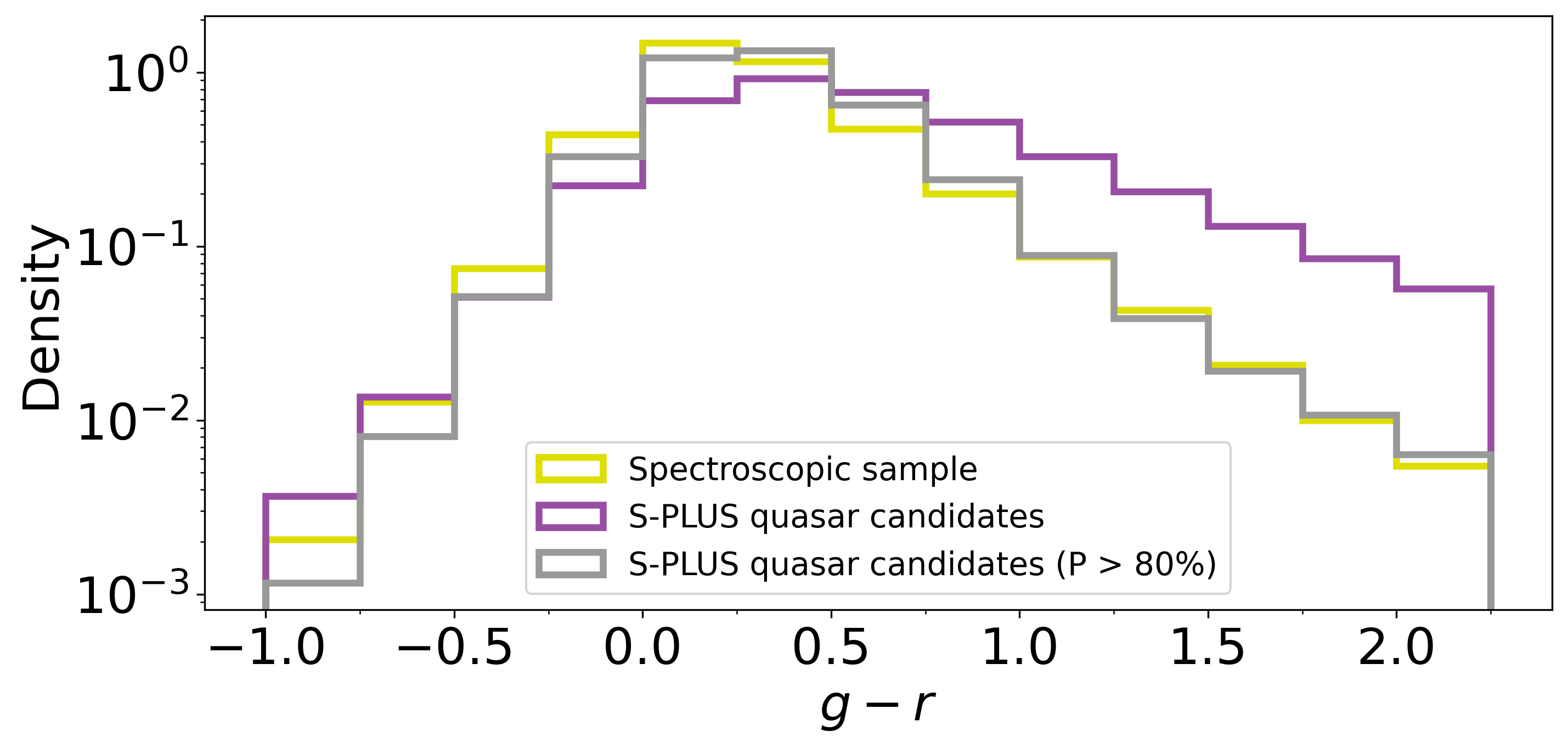}}
    \label{fig:train_test_b}
\caption{Distributions of (a) magnitude $r$ and (b) colour $g-r$ for the full spectroscopic sample (yellow), all quasar candidates in S-PLUS DR4 (purple), and high-confidence quasar candidates with probabilities down to 80\% (grey). Sample sizes are 33\,075, 15\,908\,524, and 1\,078\,149, respectively. The $g-r$ histogram comprises 99.2\%, 93.1\%, and 99.6\% of each sample, respectively, in the plotted range.}
\label{fig:spec_phot_dist}
\end{figure}

\begin{figure*}

\centering
\subfloat[]{\includegraphics[width=0.30\textwidth]{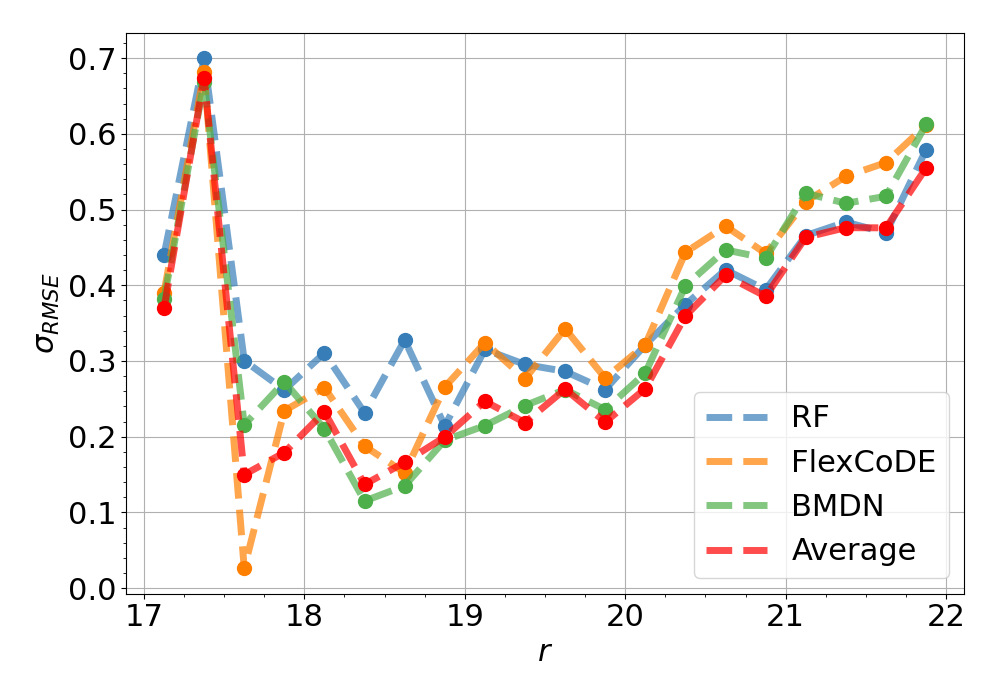}
\label{all:a}}
\hfil
\subfloat[]{\includegraphics[width=0.30\textwidth]{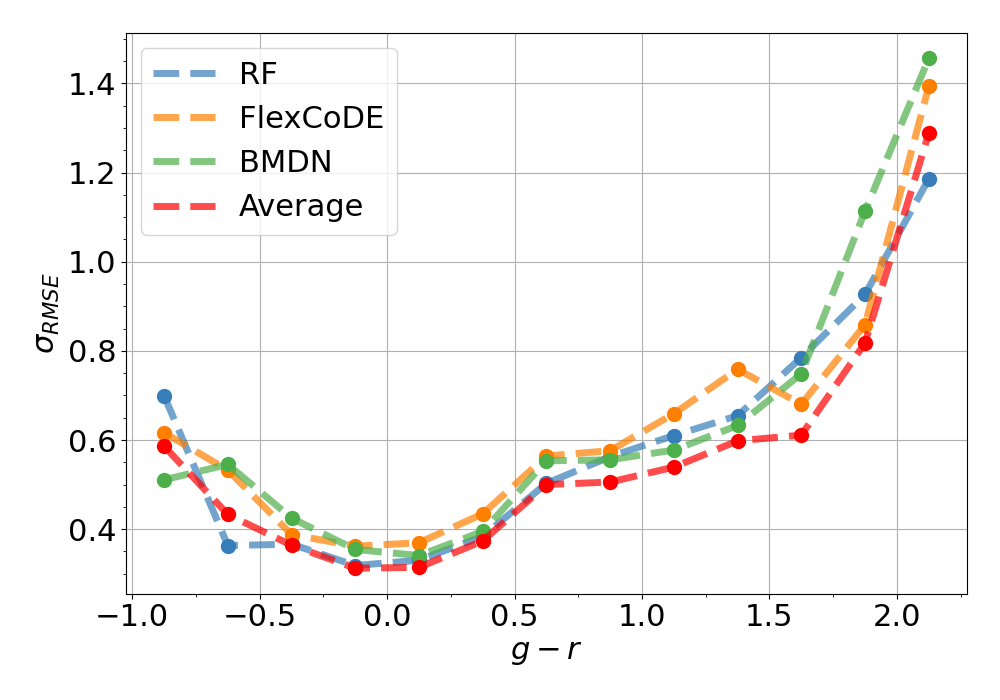}%
\label{all:b}}
\hfil
\subfloat[]{\includegraphics[width=0.30\textwidth]{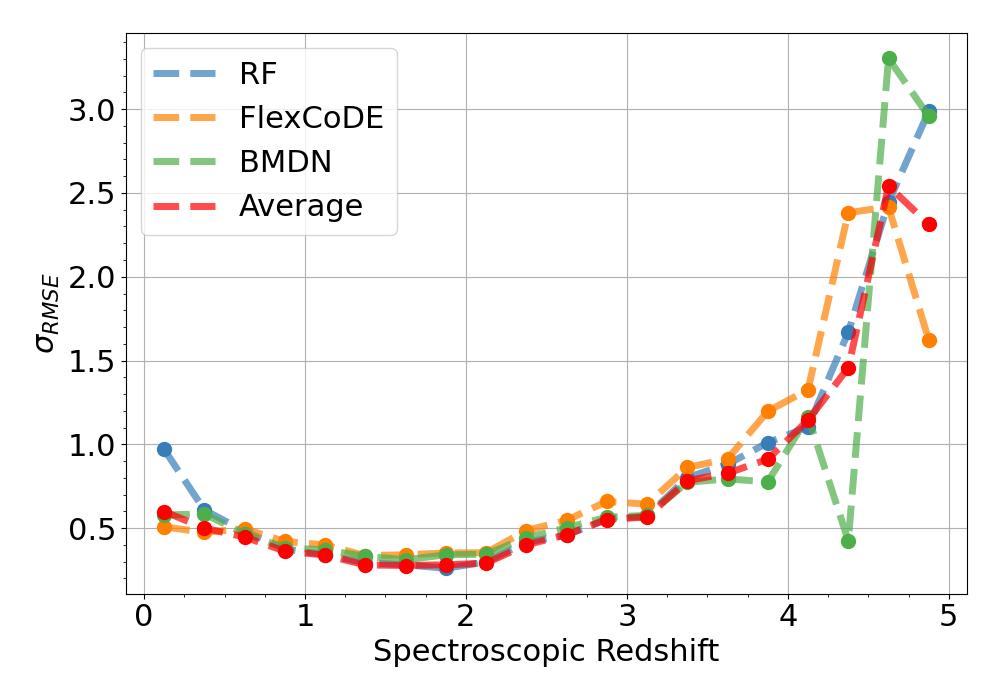}%
\label{all:c}}
\hfil

\centering
\subfloat[]{\includegraphics[width=0.30\textwidth]{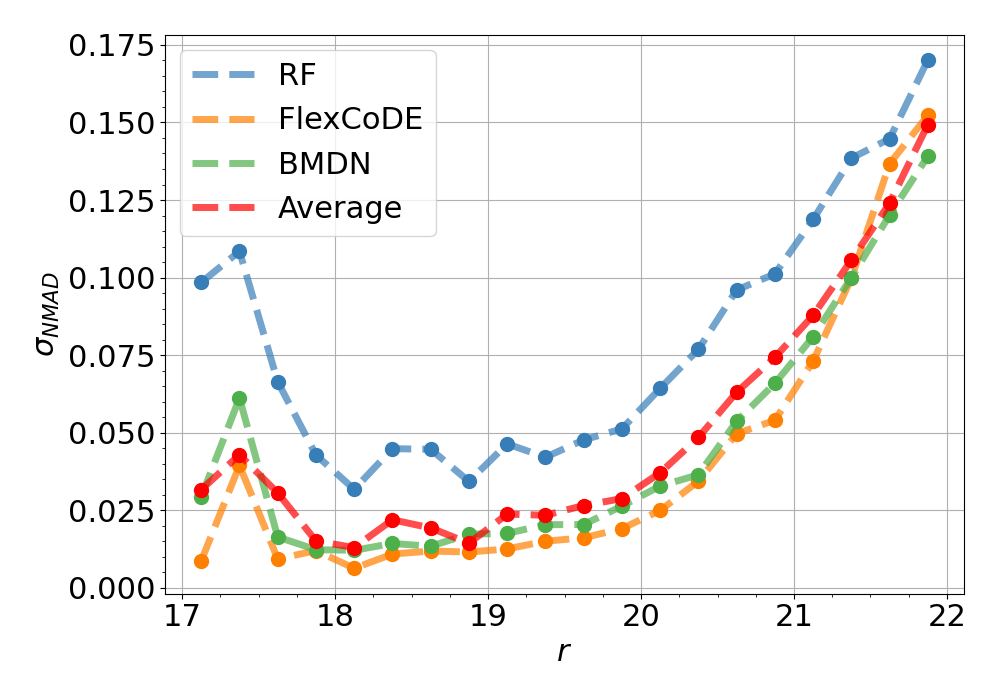}
\label{all:d}}
\hfil
\subfloat[]{\includegraphics[width=0.30\textwidth]{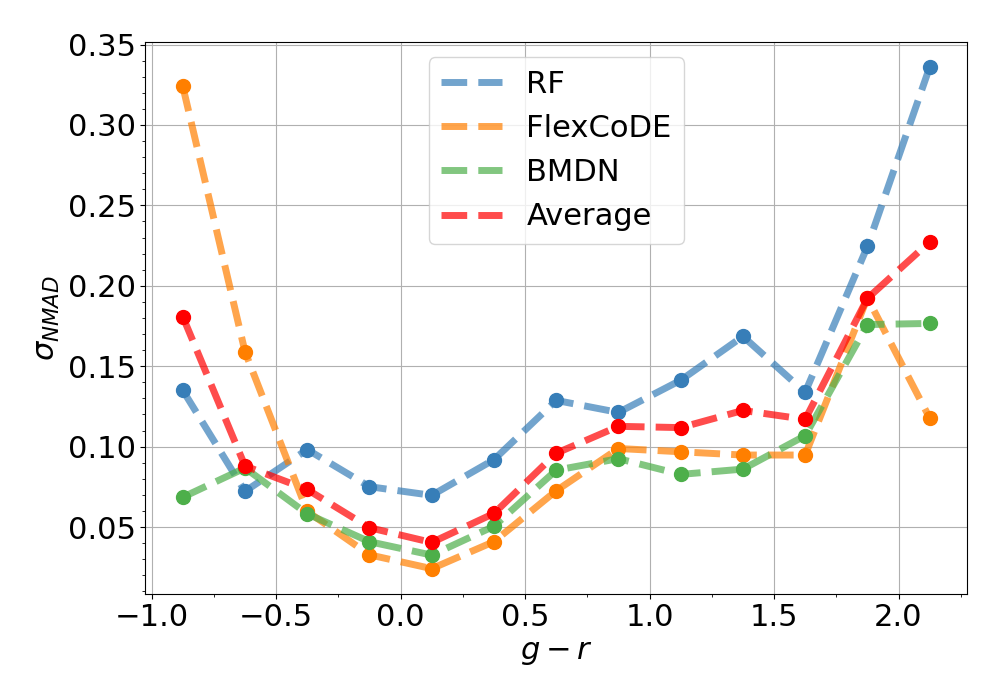}%
\label{all:e}}
\hfil
\subfloat[]{\includegraphics[width=0.30\textwidth]{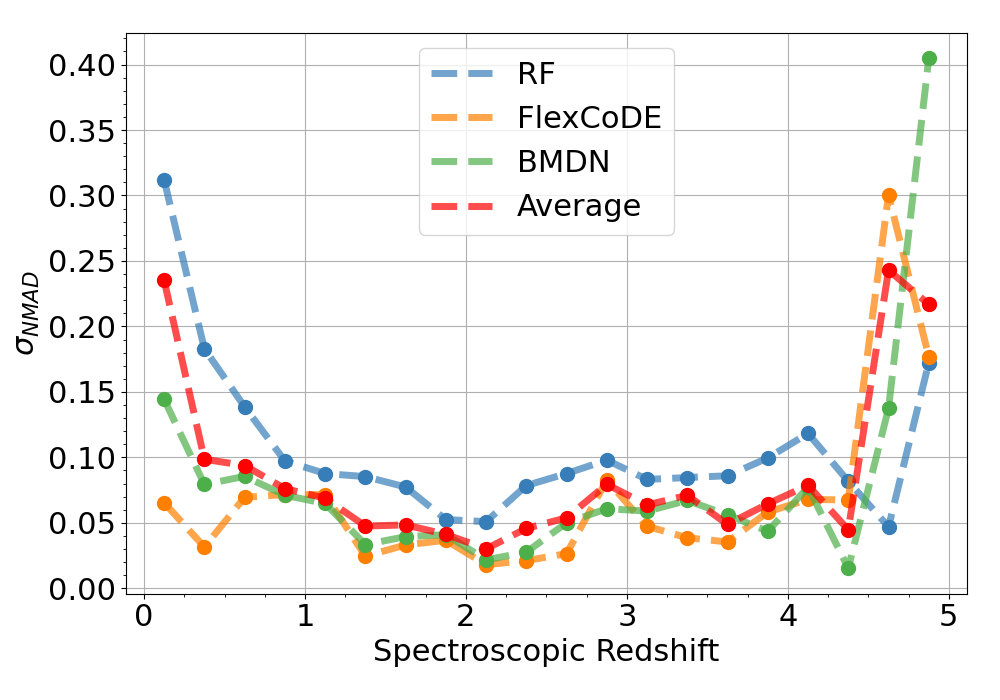}%
\label{all:f}}
\hfil

\centering
\subfloat[]{\includegraphics[width=0.30\textwidth]{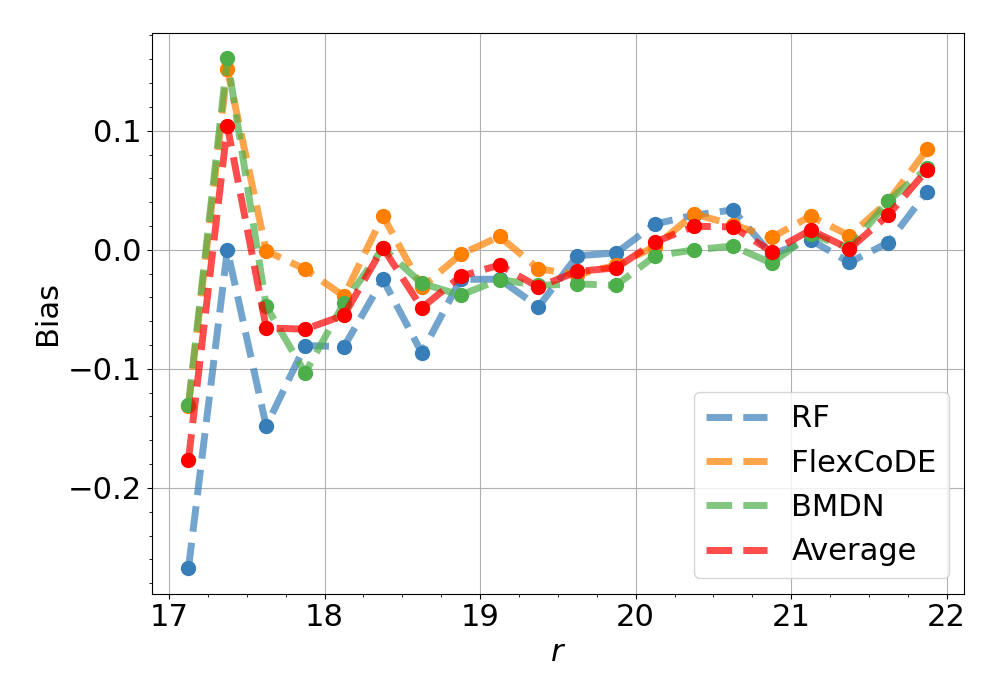}%
\label{all:g}}
\hfil
\subfloat[]{\includegraphics[width=0.30\textwidth]{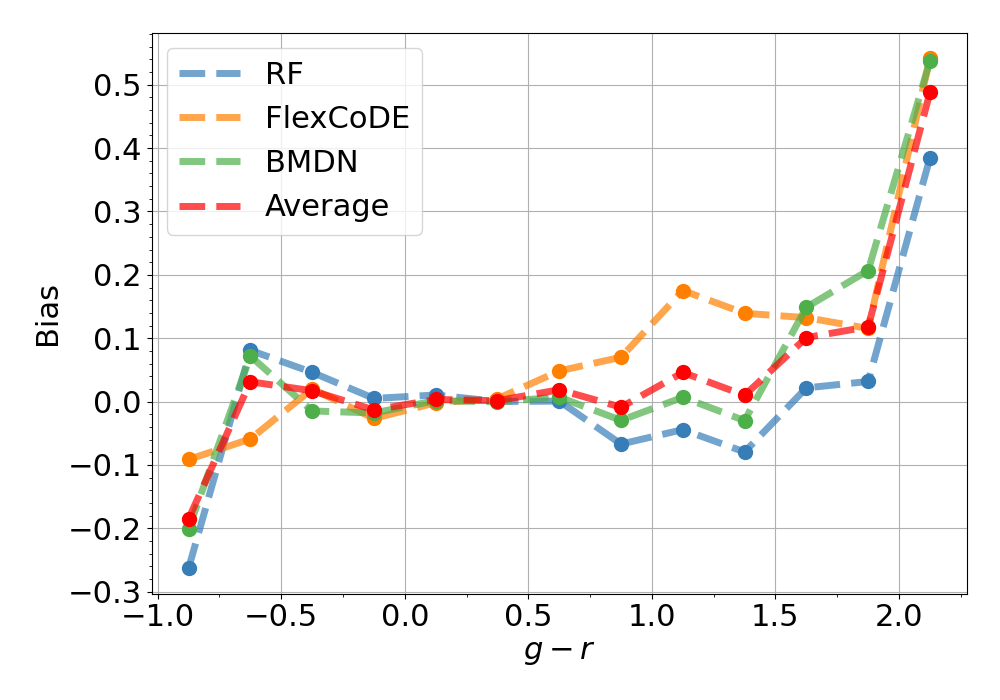}%
\label{pre:h}}
\hfil
\subfloat[]{\includegraphics[width=0.30\textwidth]{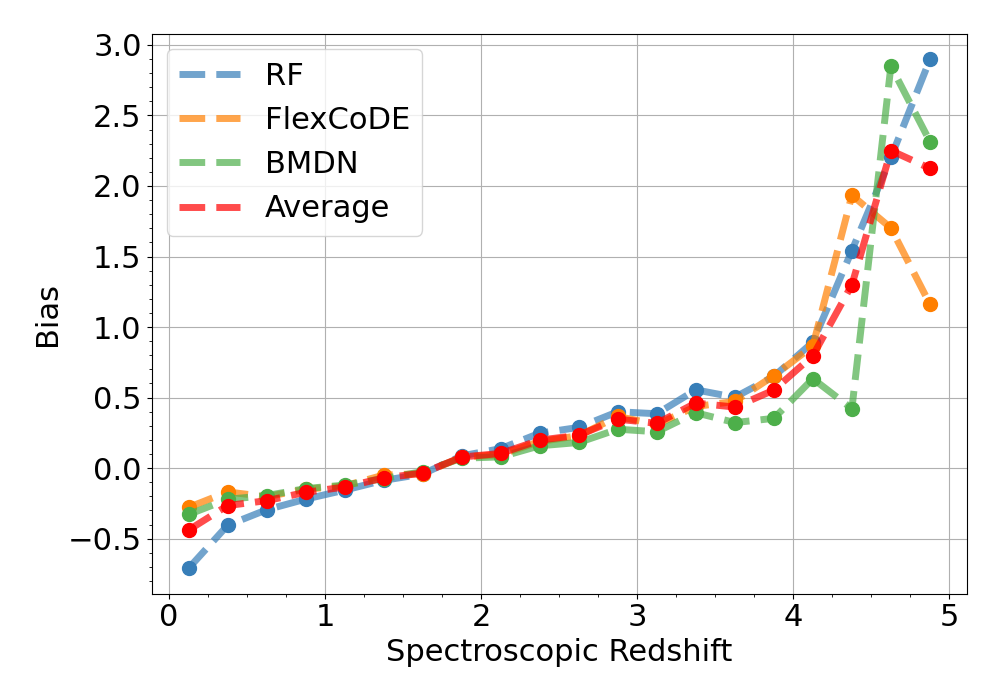}%
\label{all:i}}
\hfil

\centering
\subfloat[]{\includegraphics[width=0.30\textwidth]{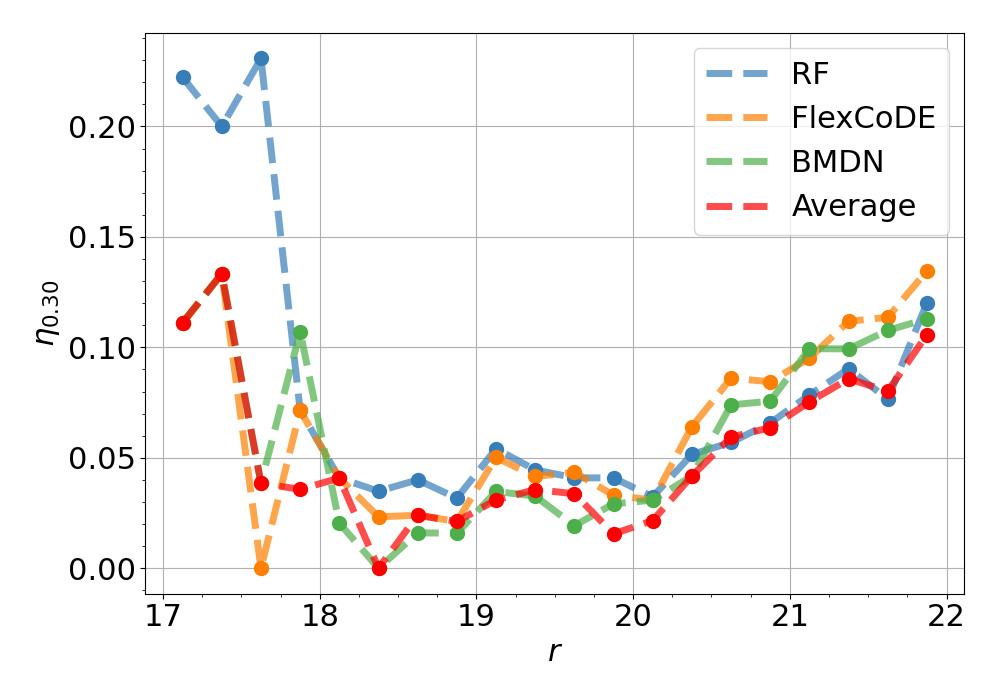}%
\label{all:j}}
\hfil
\subfloat[]{\includegraphics[width=0.30\textwidth]{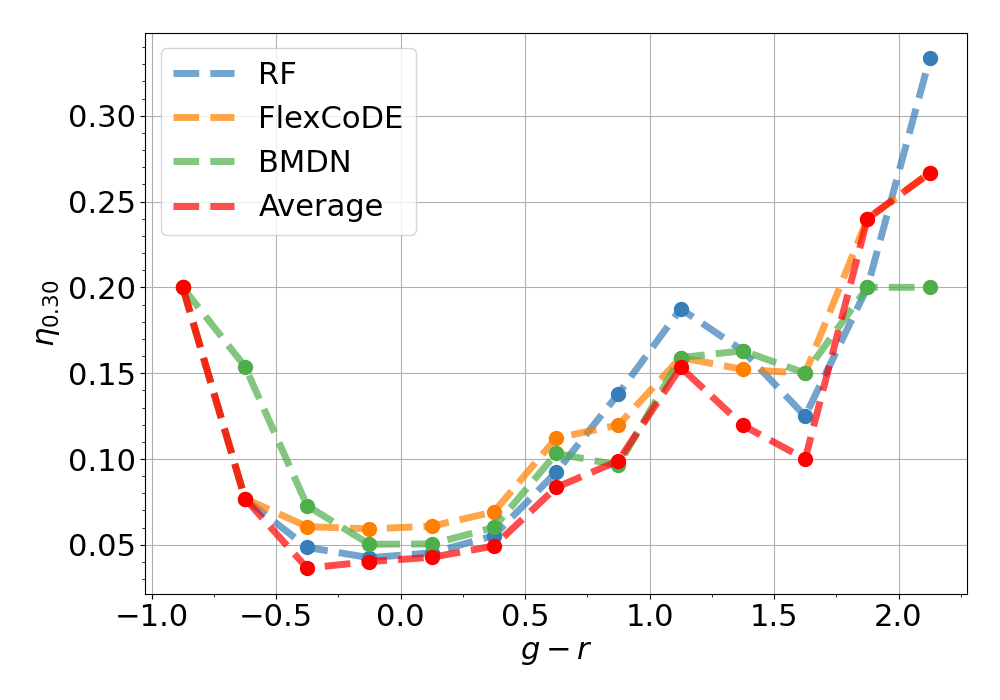}%
\label{all:k}}
\hfil
\subfloat[]{\includegraphics[width=0.30\textwidth]{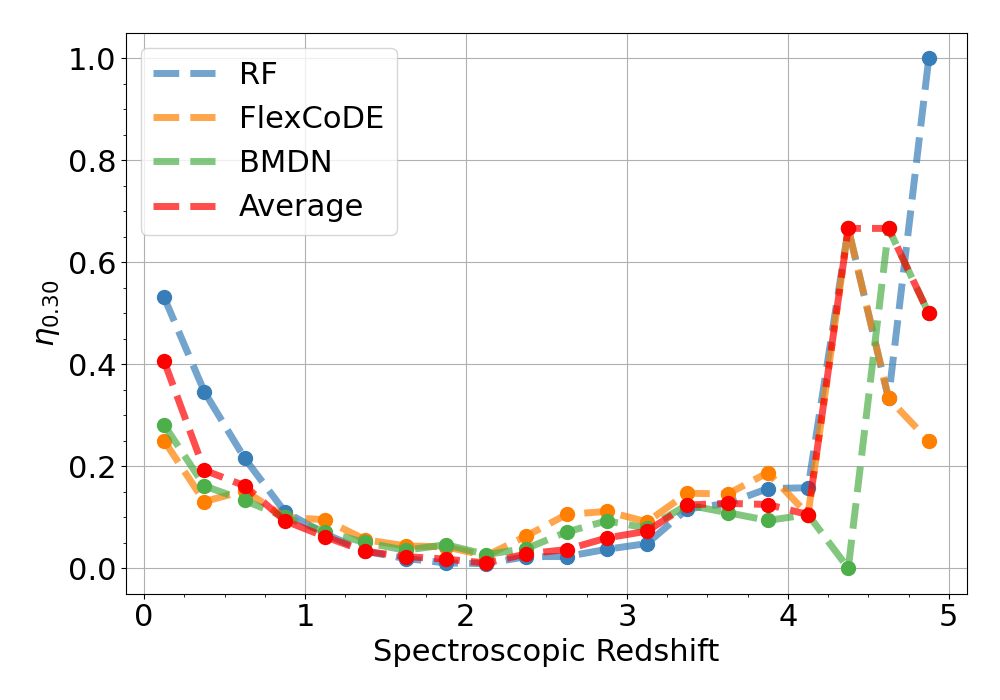}%
\label{all:l}}
\hfil

\caption{Performances in terms of $\sigma_{\text{RMSE}}$ (panels a, b, c), $\sigma_{\text{NMAD}}$ (panels d, e, f) , $\mu$ (panels g, h, i), and $\eta_{0.30}$ (panels j, k, l) per bin of magnitude r (left panels), colour $g-r$ (middle panels), and spectroscopic redshift (right panels). The curves shown in the plots refer to the Random Forest, FlexCoDE, and Bayesian Mixture Density Network trained with the \texttt{broad+GALEX+WISE+narrow} colour space, as well as the average of the predictions from the three methods.}
\vspace{-10 pt}  
\label{fig:all_models}
\end{figure*}

Confirming which method is generalising better for unseen data is not trivial due to the intrinsic biases of using spectroscopic datasets to train the models. We advise users to use the information we provide in the catalogue with extra caution for the candidates that extrapolate the colour space distribution of the spectroscopic sample (for instance, see Fig. \ref{fig:spec_phot_dist}). Although we recommend \flex{} as it presented the lowest CDE loss, some science cases will demand certain requirements over specific magnitude/redshift ranges. For those, the performances per bin of magnitude $r$, colour $g-r$ and spectroscopic redshift allow for a more precise decision on which method to use (Fig. \ref{fig:all_models}). Note that performances are poorer for lower ($z\lesssim0.5$) and higher ($z\gtrsim4.2$) probably due to the lack of objects in this range (see Fig. \ref{fig:train_test_distribution}). Performances for brighter  ($r\lesssim17.5$) or bluer/redder ($|g-r|\gtrsim0.5$) objects are also affected for the same reason (see Fig. \ref{fig:spec_phot_dist}). The performance decrease as magnitude increases is most likely related to increasing photometric uncertainties. 

\raggedbottom	


\section{Discussion and Conclusions}
\label{sec:discussion}

Up to this point, S-PLUS is the Southern hemisphere survey that provided data within the largest area ($\sim$3000 square degrees in \dr) and the highest number of narrow bands. This makes S-PLUS a good laboratory for empirically evaluating the narrow-band importance in tasks such as photometric redshift (photo-\textit{z}) regression. In this work, we compared three independent methods (Random Forest, \flex{}, Bayesian Mixture Density Network) trained with S-PLUS, GALEX, and WISE colours to obtain quasar photo-\textit{z}s. 

A reasonable improvement in quasar photo-\textit{z} predictions was observed when narrow-band colours were included in the model over training with only broad-band colours. On the other hand, narrow-band information provided a small impact on the single-point estimates when GALEX and WISE colours were available. Thus narrow bands play a more important role for objects that do not have GALEX or WISE counterparts, which respectively makes 92\% and 25\% of the S-PLUS data. This conclusion is consistent with all tested algorithms. 

As multimodal photo-\textit{z} distributions are expected due to the colour-redshift degeneracy, it is more appropriate to use the whole PDF information instead of single-point estimate summaries when evaluating performances. The models trained with narrow-band provide sharper distributions than the broad-band models, suggesting that the prediction uncertainties are not being properly accounted. This is confirmed through the PIT distributions, which show that the estimated PDFs are underdispersed for both \flex{} and \bnn{} (in this work, we do not analyse \rf{}'s PDFs) and a recalibration procedure is still needed. Nevertheless, for the scope of this paper, we have good indications that narrow-band photometry is helping to provide more accurate photo-\textit{z}s, based on the CDE loss (which accounts for the PDF and not only the single-point estimates) and feature importance estimates for the tree-based models (especially from \flex{}).

Within this paper, we provide a value-added catalogue of photometric redshifts from all tested methods for 1414 fields ($\sim$3\,000 $\text{deg}^2$) of S-PLUS DR4. We broadly recommend using the narrow-band \flex{} model as it presents the best CDE loss.

We conclude with a comment about the next-generation survey, the Rubin Observatory's LSST (\citealt{2019ApJ...873..111I}). LSST
will obtain broad-band photometry for billions of galaxies and millions of quasars. 
Given our analysis here, photometric redshift performance for these objects could be improved in two principal ways. First, the wavelength coverage could be extended over that
accessible to Rubin Observatory using near-infrared
space-based survey facilities such as Euclid \citep{2011arXiv1110.3193L} and Roman Space Telescope \citep{2015arXiv150303757S}.
Alternatively, narrow-band photometry could be obtained after the initial 10-year broad-band survey is completed using the same telescope and camera as for LSST, but with an updated filter set as
discussed in e.g. \cite{2019BAAS...51c.303Y} and \cite{2019BAAS...51g.273K}.

\raggedbottom

\section*{Acknowledgments}

LN acknowledges Alberto Krone-Martins, Kai Polsterer, and Pedro Bernadinelli for thoughtful scientific discussions and Lara Caldas, for giving the great acronym suggestion. The authors thank the referee for the suggestions that leveraged the discussions presented in this work. The authors also thank Eduardo Telles, Stephane Werner, and Paulo Afranio Lopes for reviewing this manuscript.

This work was supported by a PhD fellowship to the lead author from Funda\c{c}\~{a}o de Amparo \`{a} Pesquisa do Estado de S\~{a}o Paulo (FAPESP) with grants 2019/01312-2 and 2021/12744-0. 
RRV acknowledges the support from CNPq (grant 2020-538), FAPESP (grants 2021/08983-0 and 2023/05003-0) and CAPES (grant 88887.821818/2023-00). 
RI is grateful for the financial support of FAPESP (grants 2019/11321-9 and 2023/07068-1) and the Brazilian National Research Council (CNPq; grants 309607/2020-5 and 422705/2021-7).
EVRL acknowledges the financial support of the Coordination for the Improvement of Higher Education Personnel 
 (CAPES; grant 88887.470064/2019-00). 
NSTH acknowledges the support from FAPESP (2015/22308-2 and 2022/15304-4). 
LSJ acknowledges the support from CNPq (308994/2021-3)  and FAPESP (2011/51680-6). 
FA-F acknowledges funding for this work from FAPESP grants 2018/20977-2 and 2021/09468-1. CMdO thanks support from FAPESP grant 2019/26492-3 and CNPq grant number 309209/2019-6. 

The S-PLUS project, including the T80-South robotic telescope and the S-PLUS scientific survey, was founded as a partnership between FAPESP, the Observatório Nacional (ON), the Federal University of Sergipe (UFS), and the Federal University of Santa Catarina (UFSC), with important financial and practical contributions from other collaborating institutes in Brazil, Chile (Universidad de La Serena), and Spain (Centro de Estudios de Física del Cosmos de Aragón, CEFCA). We further acknowledge financial support from the Fundação de Amparo à Pesquisa do Estado do RS (FAPERGS), CNPq, CAPES, the Carlos Chagas Filho Rio de Janeiro State Research Foundation (FAPERJ), and the Brazilian Innovation Agency (FINEP). The authors who are members of the S-PLUS collaboration are grateful for the contributions from CTIO staff in helping in the construction, commissioning and maintenance of the T80-South telescope and camera. We are also indebted to Rene Laporte and INPE, as well as Keith Taylor, for their important contributions to the project. From CEFCA, we particularly would like to thank Antonio Marín-Franch for his invaluable contributions in the early phases of the project, David Cristóbal-Hornillos and his team for their help with the installation of the data reduction package jype version 0.9.9, César Íñiguez for providing 2D measurements of the filter transmissions, and all other staff members for their support with various aspects of the project.


\section*{Data Availability}

The data underlying the machine learning training processes of this article are available in a GitHub repository at \url{https://github.com/marixko/qucats_paper/tree/main/_survey/crossvalidation}. Data products from this article are publicly available for queries at \url{https://splus.cloud/}.

\raggedbottom



\bibliographystyle{mnras}
\bibliography{qso} 






\bsp	
\label{lastpage}
\end{document}